\documentclass[usegraphicx,usenatbib]{mn2e}

\usepackage[total={17.8cm,24.0cm},centering]{geometry}
\usepackage{times}

\newcommand{\apj}{ApJ}           
\newcommand{\mnras}{MNRAS}       
\newcommand{\sauron}{\texttt{SAURON}}
\newcommand{\kms}{\hbox{km s$^{-1}$}}
\newcommand{\plotone}[1]{\includegraphics[width=\columnwidth]{#1}}
\newcommand{\refsec}[1]{Section~\ref{#1}}
\newcommand{\reffig}[1]{Fig.~\ref{#1}}

\title[The SAURON project -- IV]
{The SAURON project -- IV. The mass-to-light ratio, \\
the virial mass estimator and the fundamental plane\\
of elliptical and lenticular galaxies}

\author[Cappellari et al.]{Michele Cappellari,$^1$\thanks{E-mail:
cappellari@strw.leidenuniv.nl}
R.\ Bacon,$^2$
M.\ Bureau,$^3$
M.~C.\ Damen,$^1$
Roger L.\ Davies,$^3$
\newauthor
P.~Tim de Zeeuw,$^1$
Eric Emsellem,$^2$
Jes\'us Falc\'on-Barroso,$^1$
Davor Krajnovi\'c,$^3$
\newauthor
Harald Kuntschner,$^4$
Richard M.\ McDermid,$^1$
Reynier F.\ Peletier,$^5$
Marc Sarzi,$^3$
\newauthor
Remco C.~E.\ van den Bosch,$^1$
and Glenn van de Ven$^1$\\
$^1$Leiden Observatory, Postbus 9513, 2300 RA Leiden, The Netherlands\\
$^2$Centre de Recherche Astrophysique de Lyon, 9~Avenue Charles
    Andr\'e, 69230 Saint Genis Laval, France\\
$^3$Denys Wilkinson Building, University of Oxford, Keble Road,
    Oxford, United Kingdom \\
$^4$Space Telescope European Coordinating Facility, European Southern
    Observatory, Karl-Schwarzschild-Str~2, 85748 Garching, Germany\\
$^5$Kapteyn Astronomical Institute, Postbus 800, 9700 AV Groningen,
    The Netherlands}

\pagerange{\pageref{firstpage}--\pageref{lastpage}} \pubyear{2006}

\begin{document}
\label{firstpage}
\maketitle
%
%
\begin{abstract}
We investigate the well-known correlations between the dynamical mass-to-light ratio $M/L$ and other global observables of elliptical (E) and lenticular (S0) galaxies. We construct two-integral Jeans and three-integral Schwarzschild dynamical models for a sample of 25 E/S0 galaxies with \sauron\ integral-field stellar kinematics to about one effective (half-light) radius $R_{\rm e}$. They have well-calibrated $I$-band \emph{Hubble Space Telescope} WFPC2 and large-field ground-based photometry, accurate surface brightness fluctuation distances, and their observed kinematics is consistent with an axisymmetric intrinsic shape. All these factors result in an unprecedented accuracy in the $M/L$ measurements. We find a tight correlation of the form $(M/L)=(3.80\pm0.14) \times (\sigma_{\rm e}/200\;\kms)^{0.84\pm0.07}$ between the $M/L$ (in the I-band) measured from the dynamical models and the luminosity-weighted second moment $\sigma_{\rm e}$ of the line-of-sight velocity-distribution within $R_{\rm e}$. The observed rms scatter in $M/L$ for our sample is 18\%, while the inferred intrinsic scatter is $\sim13\%$.  The $(M/L)$--$\sigma_{\rm e}$ relation can be included in the remarkable series of tight correlations between $\sigma_{\rm e}$ and other galaxy global observables.  The comparison of the observed correlations with the predictions of the Fundamental Plane (FP), and with simple virial estimates, shows that the `tilt' of the FP of early-type galaxies, describing the deviation of the FP from the virial relation, is almost exclusively due to a real $M/L$ variation, while structural and orbital non-homology have a negligible effect.  When the photometric parameters are determined in the `classic' way, using growth curves, and the $\sigma_{\rm e}$ is measured in a large aperture, the virial mass appears to be a reliable estimator of the mass in the central regions of galaxies, and can be safely used where more `expensive' models are not feasible (e.g.\ in high redshift studies). In this case the best-fitting virial relation has the form $(M/L)_{\rm vir}=(5.0\pm0.1)\times R_{\rm e}\sigma_{\rm e}^2/(L\, G)$, in reasonable agreement with simple theoretical predictions. We find no difference between the $M/L$ of the galaxies in clusters and in the field. The comparison of the dynamical $M/L$ with the $(M/L)_{\rm pop}$ inferred from the analysis of the stellar population, indicates a median dark matter fraction in early-type galaxies of $\sim30\%$ of the total mass inside one $R_{\rm e}$, in broad agreement with previous studies, and it also shows that the stellar initial mass function varies little among different galaxies. Our results suggest a variation in $M/L$ at constant $(M/L)_{\rm pop}$, which seems to be linked to the galaxy dynamics. We speculate that fast rotating galaxies have lower dark matter fractions than the slow rotating and generally more massive ones. If correct, this would suggest a connection between the galaxy assembly history and the dark matter halo structure.  The tightness of our correlation provides some evidence against cuspy nuclear dark matter profiles in galaxies.
\end{abstract}
%
%
\section{Introduction}
\label{sec:intro}

Early-type galaxies display a large variety of morphologies and kinematics. Some of them show little rotation, significant kinematic twists, and kinematically-decoupled components in their centres. These systems are generally classified as ellipticals (E) from photometry alone, and are more common among the most massive objects, which also tend to be redder, metal-rich and to appear nearly round on the sky. The remaining early-type galaxies show a well-defined rotation pattern, with a rotation axis generally well aligned with the photometric minor axis. These objects are often less massive, tend to be bluer and can appear very flat on the sky. These galaxies are generally classified either as E or as lenticulars (S0) from photometry \citep{dav83,fra91,kor96}.

Despite the differences between individual objects, when considering early-type galaxies as a class, it appears that they satisfy a number of regular relations between their global parameters. The first to be discovered was the relation $L\propto\sigma^4$ between the luminosity and the velocity dispersion of the stars in elliptical galaxies \citep{fab76}. These authors also already realized that, under simple assumptions, the observed relation implies a variation of the mass-to-light ratio ($M/L$) with galaxy luminosity.

It is now clear that the Faber-Jackson relation is the projection of a thin plane, the Fundamental Plane \citep[FP;][]{dre87,djo87}, which correlates three global observables: the effective radius $R_{\rm e}$, the velocity dispersion $\sigma$ and the surface brightness $I_e$ at the effective radius. The relation is of the form $R_{\rm e}\propto\sigma^\alpha I_{\rm e}^\beta$.  If galaxies were homologous stellar systems in virial equilibrium, with constant $M/L$, one would expect a correlation of the form $R_{\rm e}\propto\sigma^2 I_{\rm e}^{-1}$. The observed relation has instead the form $R_{\rm e}\propto\sigma^{1.24} I_{\rm e}^{-0.82}$ \citep*{jor96}, with generally good agreement between different estimates and a weak dependence on the photometric band in the visual region \citep[and references therein]{ber03,col01}.

One way to explain the observed deviation, or `tilt', of the FP from the virial predictions is to assume a power-law dependence of the $M/L$ on the other galaxy structural parameters (mainly galaxy mass). The $M/L$ variation could be due to differences in the stellar population or in the dark matter fraction in different galaxies. Other options are also possible however. The virial prediction for the FP is based on the assumption that galaxies form a set of homologous systems, both in the sense of having self-similar density distributions and in terms of having the same orbital distribution. These spatial and dynamical non-homologies could then also explain the observed tilt of the FP.  All these possible effects are known to occur in practice, and they must contribute, to some degree, to the observed tilt of the FP \citep[e.g.][]{van95}.

Various attempts have been made to estimate the contribution of the possible effects, with sometimes contradictory results.  A number of authors tried to study the origin of the FP tilt using approximate models. \citet*{cio96} used theoretical arguments to show that non-homology, particularly in the surface-brightness profile of galaxies, could in principle play a major role in the tilt. Subsequent studies concluded that non-homology is indeed significant from the indirect argument that the expected change in $M/L$ of the stellar population alone cannot explain the observed $M/L$ variation \citep{pru96,pah98,for98,for99}. From the observed spatial non-homology, namely the inverse correlation of galaxy concentration and luminosity \citep{cao93,gra01} various works concluded that the contribution of the spatial non-homology of galaxies of different luminosities is responsible for a significant fraction of the observed FP tilt \citep{pru97,gra97,tru04}. These authors constructed simple spherical isotropic models and used large samples of galaxies to estimate the variation in the virial $M/L$ from the variation of the shape of galaxy profiles with luminosity. Although changes in the light-profile shape do not, on their own, significantly affect the tilt of the FP \citep[e.g.][]{tru01} they produce a variation in the velocity dispersion profile which may influence the measured $M/L$.

\citet{vdm91}, \citet{mag98} and \citet{ger01} constructed detailed dynamical models of smaller samples of galaxies, reproducing in detail the photometry and long-slit spectroscopy observations. They also investigated the $M/L$ and concluded that the variations observed in the models are consistent with the observed FP tilt. However, the fact that the spread in their measured $M/L$ values was much larger than that in the FP, made the results inconclusive.

In this work we revisit what was done in the latter studies, with some crucial differences:
\begin{enumerate}
	\item The quality of our data allows a dramatic improvement in the accuracy of the dynamical $M/L$ determinations; for our galaxy sample \sauron\ \citep[Paper~I]{bac01} integral-field observations of the stellar kinematics are available, together with HST/WFPC2 and ground-based MDM photometry in the I-band (which we reproduce in detail with our models including ellipticity variations), and reliable distances based on the surface brightness fluctuations (SBF) determinations by \citet{ton01};

	\item Our sample was extracted from the \sauron\ representative sample of \citet[hereafter Paper~II]{dez02}, which spans a wide range in velocity dispersion and luminosity, and includes both bright and low-luminosity E and S0 galaxies;

	\item In contrast to previous studies the axisymmetric modelling technique we use can also be applied to very flattened galaxies, or objects containing multiple photometric components.
\end{enumerate}
All this allows us to carefully estimate the intrinsic scatter in the correlations between the dynamical $M/L$ and other galaxy observables, and makes it possible to set tight constraints on the measured slopes, for stringent comparisons with previous FP results and predictions of galaxy formation theory.

The paper is organized as follows. In Section~2 we present the sample selection and the photometric and kinematical data used. In Section~3 we describe the two-integral and three-integral dynamical models used to measure the $M/L$. In Section~4 we study the correlation of the $M/L$ with global galaxy parameters. We discuss our results in Section~5 and summarize our conclusions in Section~6.

\section{Sample and Data}
\label{sample}

\subsection{Selection}
\label{selection}

The set of galaxies we use for this work was extracted from the \sauron\ sample of 48 E and S0 galaxies \citep[classification taken from][hereafter RC3]{dev91}, which is representative of nearby bright early-type galaxies ($cz\le3000$ \kms; $M_B\le-18$ mag). As described in Paper~II, it contains 24 galaxies in each of the E and S0 subclasses, equally divided between `cluster' and `field' objects (the former defined as belonging to the Virgo cluster, the Coma I cloud, or the Leo I group, and the latter being galaxies outside these clusters), uniformly covering the plane of ellipticity, $\varepsilon$, versus absolute blue magnitude, $M_B$.

One of the requirements for the subsample defined in the present paper was the availability of an accurate distance determined using the SBF technique by \citet{ton01}. 38 objects in the \sauron\ sample satisfy this criterion. The second requirement was the availability of \emph{Hubble Space Telescope} (HST) photometry obtained with the WFPC2 in the $I$-band (F814W). We selected this band because it is minimally affected by intrinsic dust effects and it is expected to trace well the bulk of the luminous mass in galaxies. 35 galaxies satisfy this second criterion. The intersection between the two groups with SBF distances and $I$-band photometry leads to a sample of 29 galaxies. Out of these galaxies we additionally eliminated the objects showing strong evidence of bars within the \sauron\ field, from either the photometry (see \refsec{sec:mge}) or the kinematics. Specifically we excluded the five galaxies NGC~1023, NGC~2768, NGC~3384, NGC~3489 and NGC~4382. We did not attempt to eliminate all possibly barred galaxies from our sample, but at least we excluded the objects for which it makes little sense to deproject the observed surface density under the assumption of axisymmetry and to fit an axisymmetric model, which has to produce a bi-symmetric velocity field aligned with the photometry. We added the special galaxy M32, which is not part of the representative sample, to the remaining sample of 24 galaxies. This last object allows us to sample the low-luminosity end of our correlations, which by construction cannot be accessed with galaxies from the main survey. This leads to a final sample of 25 galaxies.

\subsection{Photometry}
\label{sec:phot}

Our photometric data consists of the HST/WFPC2/F814W images already mentioned, together with ground-based photometry obtained, in the same F814W filter, with the 1.3m McGraw-Hill telescope at the MDM observatory on Kitt Peak. A relatively large field-of-view of 17\farcm1$\times$17\farcm1 was used for the MDM observations to provide good coverage of the outer parts of our galaxies and to allow for an accurate sky subtraction from the observed frames. The MDM images are part of a complete photometric survey of the \sauron\ galaxies, and they will be described in detail elsewhere.

\begin{figure}
  \plotone{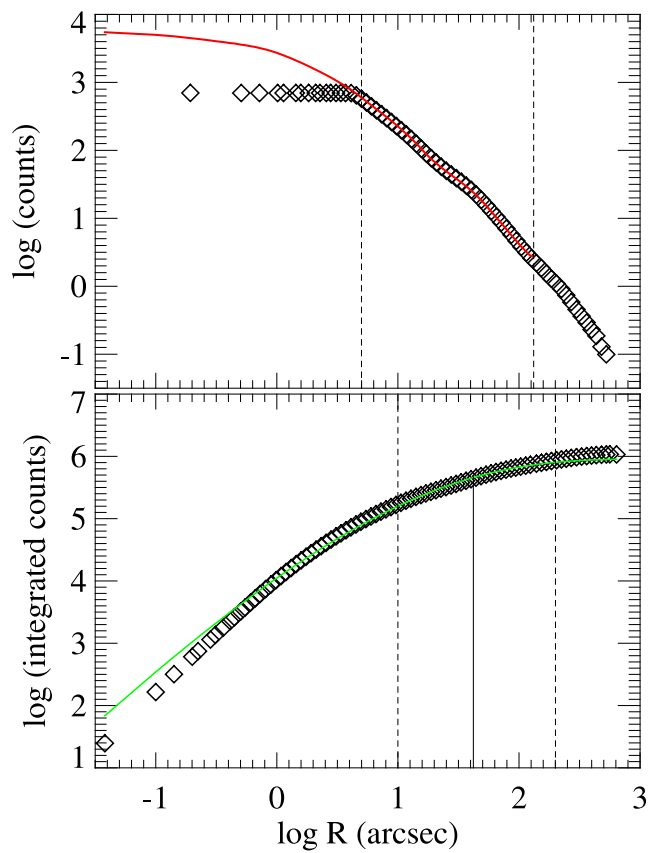}
    \caption{{\em Top Panel:} match between the photometric profile of NGC~3379 derived from the WFPC2/F814W (red solid line) and MDM images (diamonds). The sky-level of the WFPC2 image and the scaling factor of the MDM image are the free parameters of the fit. The flattening in the central ground-based profile is due to saturation of the image inside $R\la3\arcsec$. The fit was performed in the region between the two vertical dotted lines, where the WFPC2 and MDM profiles overlap, while excluding the innermost 3\arcsec\ to avoid PSF and saturation effects. {\em Bottom Panel:} fit of a de Vaucouleurs $R^{1/4}$ growth curve (green solid line) to the combined WFPC2$+$MDM observed aperture photometry (diamonds). The two vertical dotted lines define the boundaries of the region that was included in the fit, which is limited to 200\arcsec\ to reduce uncertainties due to the sky subtraction and contamination from nearby galaxies or foreground stars. The solid vertical line marks the location of $R_{\rm e}$.}
    \label{fig:profiles}
\end{figure}

We adopted the WFPC2 images as reference for the photometric calibration and  rescaled the MDM images to the same level. For this we measured logarithmically-sampled photometric profiles on the two images, using circular apertures, after masking bright stars or galaxies within the field. The sky level of the wide-field ground-based image can be determined easily and accurately by requiring the profile to tend asymptotically to a power-law at large radii. This leaves two free parameters: the scaling ratio between the ground-based image and the WFPC2 image, and the sky level of the WFPC2 image. The two parameters were fitted by minimizing the relative error between the two photometric profiles in the region of overlap, while excluding the innermost 3\arcsec\ to avoid seeing effects (top panel in \reffig{fig:profiles}). After the fit we merged the two sky-subtracted and scaled profiles into a single one, which was used for the determination of the global photometric parameters (bottom panel in \reffig{fig:profiles}).

We measured the half-light radius $R_{\rm e}$ and total galaxy luminosity as in classic studies on the FP of early-type galaxies \citep[e.g.][]{bur87,jor95}, namely from a fit of a de Vaucouleurs $R^{1/4}$ growth curve to the aperture photometry. The fit was generally restricted to radii $R\ga10\arcsec$ where the growth curve is a good description of the observations. We also experimented with alternatives: (i) fitting a \citet{ser68} $R^{1/n}$ curve to the profile, (ii) fitting a $R^{1/n}$ growth curve to the aperture photometry, (iii) fitting a $R^{1/n}$ growth curve, with the value of $n$ fixed by a previous fit to the profile. We found that the values of $R_{\rm e}$ obtained with different methods differ substantially in some cases. This is not too surprising as the determination is based on an extrapolation of the profile. We adopted the $R^{1/4}$ growth curve fit because it allows a direct comparison of our results with previous works on the FP in \refsec{sec:fp}. Another advantage  is that with a growth curve approach the profile does not have to be well described by the assumed parameterization, but one just tries to extrapolate the luminosity to infinite radius, starting from the outermost measured photometric points. This makes the measurement more robust and better reproducible than a profile-fitting approach. The measured values of the global photometric parameters are given in Table~\ref{tab1}.

\begin{table*}
\caption{Galaxy sample and measured parameters.}
\tabcolsep=3pt
\begin{tabular}{lcccccccccccc}
\hline
Galaxy Name & $R_{\rm e}$ & $R_{\rm max}/R_{\rm e}$ & $I_T\ $ & $K_T$
            &  $\sigma'_{\rm e}$ & $i$ & $(M/L)_{\rm Jeans}$
            & $(M/L)_{\rm Schw}$ & $(M/L)_{\rm pop}$ & $(m-M)$ & $A_B$ & Fast-rotator ?   \\
          & (arcsec) &      & (mag)&(mag)&(\kms)& ($^\circ$) & ($I$-band) & ($I$-band) & ($I$-band) & (mag)  & (mag) &     \\
           (1) & (2) & (3)  & (4)  & (5)  & (6) & (7)& (8)  & (9)  & (10) & (11)           & (12) & (13) \\
\hline
NGC  221 (M32) &  30 & 0.92 & 7.07 & 5.09 &  60 & 50 & 1.25 & 1.41 & 1.17 & 24.49$\pm$0.08 & 0.35 & yes \\
NGC  524       &  51 & 0.61 & 8.76 & 7.16 & 235 & 19 & 5.04 & 4.99 & 3.00 & 31.84$\pm$0.20 & 0.36 & yes \\
NGC  821       &  39 & 0.62 & 9.47 & 7.90 & 189 & 90 & 3.54 & 3.08 & 2.60 & 31.85$\pm$0.17 & 0.47 & yes \\
NGC 2974       &  24 & 1.04 & 9.43 & 8.00 & 233 & 57 & 4.47 & 4.52 & 2.34 & 31.60$\pm$0.24 & 0.23 & yes \\
NGC 3156       &  25 & 0.80 &10.97 & 9.55 &  65 & 67 & 1.70 & 1.58 & 0.62 & 31.69$\pm$0.14 & 0.15 & yes \\
NGC 3377       &  38 & 0.53 & 8.98 & 7.44 & 138 & 90 & 2.32 & 2.22 & 1.75 & 30.19$\pm$0.09 & 0.15 & yes \\
NGC 3379 (M105)&  42 & 0.67 & 8.03 & 6.27 & 201 & 90 & 3.14 & 3.36 & 3.08 & 30.06$\pm$0.11 & 0.10 & yes \\
NGC 3414       &  33 & 0.60 & 9.57 & 7.98 & 205 & 90 & 4.06 & 4.26 & 2.91 & 31.95$\pm$0.33 & 0.10 & no \\
NGC 3608       &  41 & 0.49 & 9.40 & 8.10 & 178 & 90 & 3.62 & 3.71 & 2.57 & 31.74$\pm$0.14 & 0.09 & no \\
NGC 4150       &  15 & 1.39 &10.51 & 8.99 &  77 & 52 & 1.56 & 1.30 & 0.66 & 30.63$\pm$0.24 & 0.08 & yes \\
NGC 4278       &  32 & 0.82 & 8.83 & 7.18 & 231 & 45 & 4.54 & 5.24 & 3.05 & 30.97$\pm$0.20 & 0.12 & yes \\
NGC 4374 (M84) &  71 & 0.43 & 7.69 & 6.22 & 278 & 90 & 4.16 & 4.36 & 3.08 & 31.26$\pm$0.11 & 0.17 & no \\
NGC 4458       &  27 & 0.74 &10.68 & 9.31 &  85 & 90 & 2.33 & 2.28 & 2.27 & 31.12$\pm$0.12 & 0.10 & no \\
NGC 4459       &  38 & 0.71 & 8.91 & 7.15 & 168 & 47 & 2.76 & 2.51 & 1.86 & 30.98$\pm$0.22 & 0.20 & yes \\
NGC 4473       &  27 & 0.92 & 8.94 & 7.16 & 192 & 73 & 3.26 & 2.91 & 2.88 & 30.92$\pm$0.13 & 0.12 & yes \\
NGC 4486 (M87) & 105 & 0.29 & 7.23 & 5.81 & 298 & 90 & 5.19 & 6.10 & 3.33 & 30.97$\pm$0.16 & 0.10 & no \\
NGC 4526       &  40 & 0.66 & 8.41 & 6.47 & 222 & 79 & 3.51 & 3.35 & 2.62 & 31.08$\pm$0.20 & 0.10 & yes \\
NGC 4550       &  14 & 1.45 &10.40 & 8.69 & 110 & 78 & 2.81 & 2.62 & 1.44 & 30.94$\pm$0.20 & 0.17 & yes \\
NGC 4552 (M89) &  32 & 0.63 & 8.54 & 6.73 & 252 & 90 & 4.52 & 4.74 & 3.35 & 30.87$\pm$0.14 & 0.18 & no \\
NGC 4621 (M59) &  46 & 0.56 & 8.41 & 6.75 & 211 & 90 & 3.25 & 3.03 & 3.12 & 31.25$\pm$0.20 & 0.14 & yes \\
NGC 4660       &  11 & 1.83 & 9.96 & 8.21 & 185 & 70 & 3.63 & 3.63 & 2.96 & 30.48$\pm$0.19 & 0.14 & yes \\
NGC 5813       &  52 & 0.52 & 9.12 & 7.41 & 230 & 90 & 4.19 & 4.81 & 2.97 & 32.48$\pm$0.18 & 0.25 & no \\
NGC 5845       & 4.6 & 4.45 &11.10 & 9.11 & 239 & 90 & 3.17 & 3.72 & 2.96 & 32.01$\pm$0.21 & 0.23 & yes \\
NGC 5846       &  81 & 0.29 & 8.41 & 6.93 & 238 & 90 & 4.84 & 5.30 & 3.33 & 31.92$\pm$0.20 & 0.24 & no \\
NGC 7457       &  65 & 0.39 & 9.45 & 8.19 &  78 & 64 & 1.86 & 1.78 & 1.12 & 30.55$\pm$0.21 & 0.23 & yes \\
\hline
\end{tabular}
\begin{minipage}{17.8cm}
Notes:
(1) NGC number.
(2) Effective (half-light) radius $R_{\rm e}$ measured in the $I$-band
    from HST/WFPC2 $+$ MDM images. Comparison with published values
    shows an rms scatter of 17\%.
(3) Ratio between the maximum radius $R_{\rm max}$ sampled by the kinematical
	observations and $R_{\rm e}$. We defined
	$R_{\rm max}\equiv\sqrt{A/\pi}$, where $A$ is the area on the sky sampled
	by the \sauron\ observations.
(4) Total observed $I$-band galaxy magnitude. Comparison with the
    corresponding values by \citet{ton01} indicates an
    observational error of 13\%.
(5) Total observed $K$-band galaxy magnitude, from the 2MASS Extended
    Source Catalog \citep{jar00}.
(6) Velocity dispersion derived with pPXF, by fitting a purely Gaussian LOSVD
	to the luminosity-weighted spectrum within $R_{\rm e}$. For the galaxies
	in which $R_{\rm max}/R_{\rm e}<1$ this value was determined on the whole
	\sauron\ field. Equation~(\ref{eq:sigma_r}) was used to derive the
	$\sigma_{\rm e}$ value used in our correlations, adopting as $R/R_{\rm e}$
	the value in column 3. The statistical errors on these values are
	negligible, but we adopt an error of 5\% to take systematics into account.
(7) Inclination of the best-fitting two-integral Jeans model. We do
    not attach errors as they are fully model dependent.
(8) $M/L$ of the best-fitting two-integral Jeans model.
(9) $M/L$ of the best-fitting three-integral Schwarzschild model,
     computed at the inclination of column 7. Comparison with column 8
     suggests an error of 6\% in these values.
(10) Stellar population $(M/L)_{\rm pop}$ determined from measured
	line-strength values using single stellar population models. The
	median formal error on this quantity is $\sim10\%$, but this value is
	strongly assumption dependent. See \refsec{sec:population} for details.
(11) Galaxy distance modulus of \citet{ton01}, adjusted to the Cepheid
	 zeropoint of \citet{fre01} by {\em subtracting} 0.06 mag (for a
	 discussion see Section~3.3 of \citealt{mei05}).
(12) $B$-band galactic extinction of \citet*{sch98}
     as given by the NED database. We adopt $A_I=A_B/2.22$ as in NED.
(13) Galaxy classification according to the measured value of the specific
	stellar angular momentum within one $R_{\rm e}$, from the \sauron\
	kinematics. A qualitative distinction between the two classes of galaxies
	can be seen on the kinematic maps of Paper~III. More details will
	be given in Emsellem et al.\ (in preparation).
\end{minipage}
\label{tab1}
\end{table*}

\subsection{Kinematics}
\label{sec:kinematics}

The kinematic measurements used here were presented in \citet[hereafter Paper~III]{ems04}, where further details can be found. In brief, the \sauron\ integral-field spectroscopic observations, obtained at the 4.2-m William Herschel Telescope on La Palma, were reduced and merged with the XSauron software developed at CRAL (Paper~I). They where spatially binned to a minimum signal-to-noise ratio $S/N\approx60$ using the Voronoi two-dimensional (2D) binning algorithm by \citet{cap03} and the stellar kinematics were subsequently extracted with the penalized pixel-fitting (pPXF) method of \citet{cap04a}. This provided, for each Voronoi bin, the mean velocity $V$, the velocity dispersion $\sigma$ and four higher order Gauss-Hermite moments of the velocity up to $h_3$--$h_6$ \citep{vdm93,ger93}. We estimated errors on the kinematics using 100 Monte Carlo realizations and applying the prescriptions of Section~3.4 of \citet{cap04a}. In addition we observed M32 with \sauron\ in August 2003 with the same configuration as for the other galaxies in Paper~III, and obtained two pointings resulting in a field-of-view of about 40\arcsec$\times$60\arcsec.

For each galaxy in our sample we determined $\sigma_{\rm e}$, the luminosity-weighted second moment of the line-of-sight velocity distribution (LOSVD) within the half-light radius $R_{\rm e}$. Compared to the central velocity dispersion $\sigma_{\rm c}$, which was sometimes used before, this quantity has the advantage that it is only weakly dependent on the details of the aperture used. The observed $\sigma_{\rm e}$ is an approximation to the second velocity moment which appears in the virial equation  \citep[Section~4.3]{bin87}. It is proportional to $\sqrt{M}$ (with $M$ the galaxy mass), and so is weakly dependent on the details of the orbital distribution. These are the reasons why a similar quantity (estimated however from long-slit data) was also adopted by e.g.\ \citet{geb03} to study the correlation between the mass of supermassive black holes (BHs) and the velocity dispersion of the host galaxy.

The use of integral-field observations allows us to perform the $\sigma_{\rm e}$ measurements in a more rigorous manner than was possible with long slits. We measured $\sigma_{\rm e}$ from the data by coadding all luminosity-weighted {\em spectra} within $R_{\rm e}$. The resulting single `effective spectrum' has extremely high $S/N\ga300$ per spectral element, and is equivalent to what would have been observed with a single aperture of radius $R_{\rm e}$ centred on the galaxy. We then used pPXF, with the same set of multiple templates from \citet[hereafter VZ99]{vaz99} as in Paper~III, to fit a purely Gaussian LOSVD from that spectrum to determine $\sigma$ (this time the higher order Gauss-Hermite moments are set to zero and are not fitted). This procedure takes into account in a precise manner the effect of the galaxy rotation on $\sigma_{\rm e}$, so that no extra correction is required.

\begin{figure}
  	\plotone{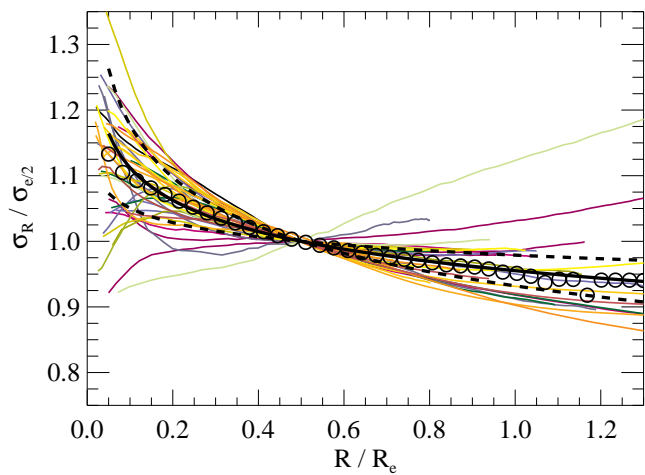}
    \caption{Luminosity-weighted second moment of the line-of-sight velocity distribution within an aperture of radius $R$, normalized to its value at $R_{\rm e}/2$ (see text details). This plot shows, with the thin coloured lines, all the 40 E/S0 galaxies in the \sauron\ sample for which $R_{\rm e}/2$ is not larger than the field-of-view. The open circles represent the biweight mean of all the curves at every given radius. The number of measurements sampled at every radius is not constant, moreover there are significant galaxy-to-galaxy variations in the profiles. The thick black line is a power-law relation $\sigma_R\propto R^{-0.066}$ which has an exponent defined by the biweight mean of the individual values for every galaxy. The dashed lines indicate the standard deviation in the individual slopes. The galaxy with the largest increase of the $\sigma_R$ profile at large radii is NGC~4550 (see Paper~III).  } \label{fig:sigma_aper}
\end{figure}

As we do not sample all galaxies out to one $R_{\rm e}$ we want to correct for this effect and to estimate how much uncertainty this can introduce in our $\sigma_{\rm e}$ measurements. For this we show in \reffig{fig:sigma_aper} the profiles of $\sigma_R$, measured by coadding the \sauron\ spectra within circular apertures of increasing radius, for all the 40 early-type galaxies in Paper~III which we sample to at least $R_{\rm e}/2$. In some cases the selected circular aperture is not fully sampled by our data, and we define the radius $R\equiv\sqrt{A/\pi}$ as that of a circular aperture with the same area $A$ on the sky actually covered by the \sauron\ data inside that aperture.
We fitted the profile of $\sigma_{\rm e}$ versus $R$ for every galaxy with a power-law relation $\sigma_R\propto R^\gamma$ and we determined the biweight \citep{hoa83} mean and standard deviation for the exponent $\gamma=-0.066\pm0.035$ in the entire set. In addition we normalized the $\sigma_R$ values to the corresponding $\sigma_{\rm e/2}$, defined as the dispersion at $R_{\rm e}/2$, and computed the biweight mean from all galaxies, at different fractions of $R_{\rm e}$. This mean $\sigma_R$ profile appears to be well described by the mean power-law relation fitted to the individual profiles. One can see that generally $\sigma_R$ decreases by less than 5\% from $R_{\rm e}/2$ up to $R_{\rm e}$, but the galaxy to galaxy variations are significant. In summary the aperture correction has the form:
\begin{equation}
	(\sigma_R/\sigma_{\rm e})=(R/R_{\rm e})^{-0.066\pm0.035}.
	\label{eq:sigma_r}
\end{equation}
The logarithmic slope of our best-fitting relation is in reasonable agreement with the value $-0.04$ found by \citet{jor95} and with the value $-0.06$ derived by \citet{meh03} from long slit spectroscopy.

The measured values of $\sigma_{\rm e}$ for our sample are given in Table~\ref{tab1} together with the actual fraction of $R_{\rm e}$ sampled by our kinematics. We decided not to restrict our $\sigma_{\rm e}$ measurements to a smaller aperture so as not to discard data for the galaxies that we sample to large radii. In what follows we correct the values of $\sigma_{\rm e}$ with the aperture correction of equation~(\ref{eq:sigma_r}) to determine our correlations and to generate our plots.

\section{Dynamical modelling}
\label{modeling}

\subsection{MGE mass model}
\label{sec:mge}

We constructed photometric models for all the 29 {\em candidate} galaxies (see beginning of \refsec{selection}) plus M32 with HST photometry and SBF distances. For this we used the Multi-Gaussian Expansion (MGE) parameterization by \citet*{ems94}, which describes the observed surface brightness as a sum of Gaussians, and allows the photometry to be reproduced in detail, including ellipticity variations and strongly non-elliptical isophotes. The accurate MGE modeling of such a large sample of galaxies, each consisting of three separate images, was made feasible by the use of the MGE fitting method and software by \citet{cap02}, which was designed with this kind of large-scale application in mind.

\begin{figure}
  	\centering
  	\includegraphics[width=0.81\columnwidth]{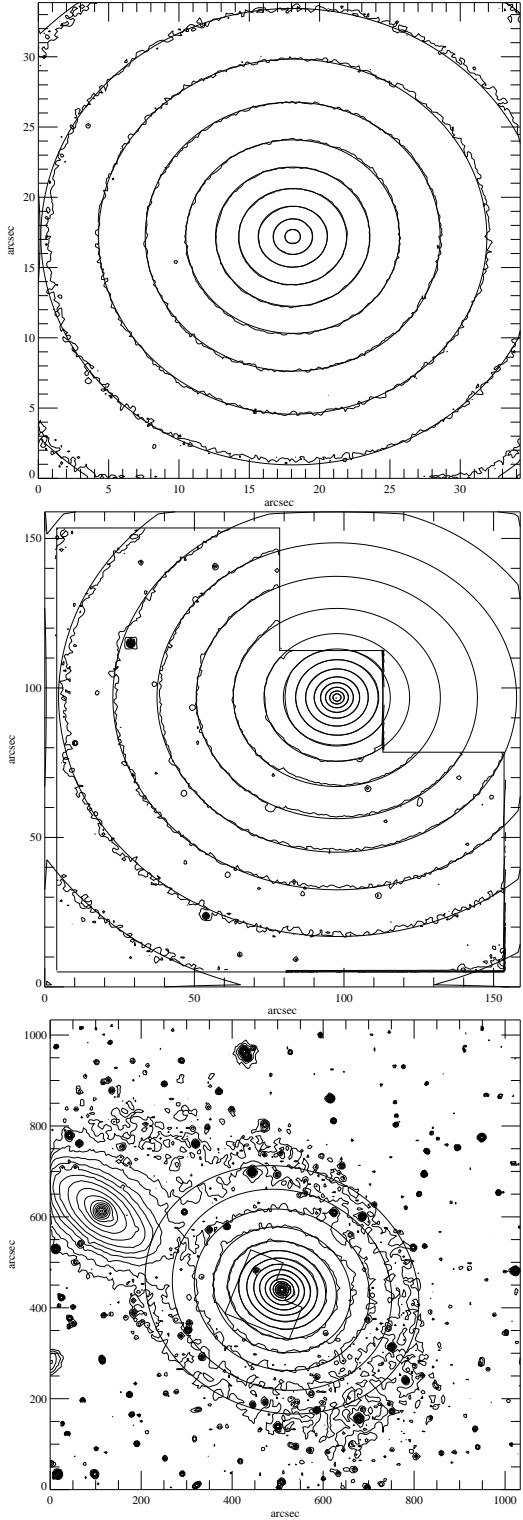}
  	\caption{Contours of the surface brightness of NGC~3379 (in 0.5 mag arcsec$^{-2}$ steps) at three different scales. {\em Top Panel:} 35\arcsec$\times$35\arcsec\ PC1 WFPC2 CCD. {\em Middle Panel:} the whole 160\arcsec$\times$160\arcsec\ WFPC2 mosaic. {\em Bottom Panel:} 17\farcm1$\times$17\farcm1\ MDM image. North is up and east is to the left. The location on the MDM field of the L-shaped WFPC2 mosaic is also shown. The galaxy at the left of the frame is NGC~3384, which was masked during the MGE fit. Superposed on the three plots are the contours of the constant-PA MGE surface brightness model, convolved with the PSF of each observation. The MGE model was fitted simultaneously to all three images.}
    \label{fig:mge_fit}
\end{figure}

Our MGE models were fitted simultaneously to three images: (i) the WFPC2/PC1 CCD (ii) the lower-resolution WFPC2 mosaic and (iii) a wide-field ground-based MDM image (\reffig{fig:mge_fit}). In \refsec{sec:phot} we described how the ground-based and WFPC2 images were carefully sky-subtracted and matched.  The MGE fits were done by keeping the position angle (PA) for all Gaussian fixed and taking PSF convolution into account \citep[our F814W MGE PSF is given in Table~3 of][]{cap02a}. The quality of the resulting fits was inspected, together with the kinematics, to exclude the 5 galaxies which could not be reasonably well fitted by a constant-PA photometric model (\refsec{selection}). This reduced the original sample to 25 galaxies.

As discussed in \citet{cap02} we `regularized' our solutions by requiring the axial ratio of the flattest Gaussian to be as big as possible, while still reproducing the observed photometry. In this way we avoided artificially constraining the possible inclinations for which the models can be deprojected assuming they are axisymmetric. This also prevents introducing sharp variations in the intrinsic density of the MGE model, unless they are required to fit the surface brightness. This regularization of the models is needed because of the non-uniqueness of the deprojection of an axisymmetric density distribution \citep{ryb87}.

The resulting analytically deconvolved MGE model for each galaxy was corrected for extinction following \citet{sch98}, as given by NED. It was converted to a surface density in solar units using the WFPC2 calibration of \citet{dol00} while assuming an $I$-band absolute magnitude for the Sun of $M_{I,\sun}=4.08$ mag \citep[all the absolute magnitudes of the Sun in this paper, in the $I$, $R$ and $K$-band, are taken from Table 2.1 of][]{bin98}. The calibrated MGE models are given in Appendix~B. The MGE model for NGC~2974 was taken from \citet{kra05}, which uses precisely the same method and calibration.

\subsection{Two-integral Jeans modeling}
\label{sec:jeans}

We constructed axisymmetric Jeans models for the 25 galaxies in our sample in the following way. Under the assumptions of the MGE method, and for a given inclination $i$, the MGE surface density can be deprojected analytically \citep{mon92} to obtain the intrinsic density $\rho(R,z)$ in the galaxy meridional plane, still expressed as a sum of Gaussians. Although the deprojection is non-unique, the MGE deprojection represents a reasonable choice, which produces realistic intrinsic densities, that resemble observed galaxies when projected from any inclination.

For an axisymmetric model with constant $M/L$, and a stellar distribution function (DF) which depends only on two integrals of motion $f=f(E,L_z)$, where $E$ is the energy and $L_z$ the angular momentum with respect to the $z$-axis (the axis of symmetry), the second velocity moments are uniquely defined \citep[e.g.][]{lyn62,hun77} and can be computed by solving the Poisson and Jeans equations \citep[e.g.][]{sat80,bin90}. The second moment $\mu_2$ projected along the line-of-sight of the model is then a function of only two free parameters, $i$ and $M/L$. We do not fit for the mass of a possible BH, as the spatial resolution of our data is not sufficient to constrain its value, but we set the mass of the central BH to that predicted by the $M_{\rm BH}$--$\sigma$ correlation \citep{fer00,geb00} as given by \citet{tre02}. The inclusion of the BH is not critical in this work, but it has a non-negligible effect for the most massive galaxies. E.g., in the case of M87 the expected BH of $\sim3\times10^9 M_{\sun}$ produces a small but detectable effect on the observed $\sigma$ up to a radius $\sim5\arcsec$ \citep[e.g.\ Fig.~13 of][]{vdm94}. To make our results virtually insensitive to the assumed BH masses we do not fit the innermost $R\la2\arcsec$ from our data, where the BH may dominate the kinematics.

Starting from the calibrated parameters of the fitted MGE model and using the MGE formalism, the model $\mu_{2,\star}$, without a BH, can be computed easily and accurately using a single one-dimensional integral via equations~(61-63)\footnote{Their equation~(63) has a typographical error and should be replaced by the following $B=\frac{1}{2}\left[ \frac{e_j^2T^4} {\sigma_j^2(1-e_j^2T^2)} + \left(\frac{e_i}{\sigma_i q_i}\right)^2\right]$} of \citet{ems94}. The contribution $\mu_{2,\bullet}$ to the second moments, due to the BH, is computed via equations (49,88,102) of \citet{ems94} and also reduces to a single one-dimensional integral. Given the linearity of the Jeans equations with respect to the potential, the luminosity weighed second moments for an MGE model with a BH can be obtained as $\nu_p\mu_2^2=\nu_p\mu_{2,\star}^2+\nu_p\mu_{2,\bullet}^2$, where $\nu_p$ is the deconvolved MGE surface brightness. The $\nu_p\mu_2^2$ has to be convolved with the PSF and integrated over the pixels before comparison with the observables.  For the 25 galaxies in our sample we computed the model predictions for $\mu_2$, at different inclinations, for each Voronoi bin on the sky. At every inclination the best fitting $M/L$ is obtained from the simple scaling relation $(M/L)\propto\mu_2^2$ as a linear least-squares fit:
\begin{equation}
	(M/L)=\left(\frac{\mathbf{d}\cdot\mathbf{d}}{\mathbf{d}\cdot\mathbf{m}}\right)^2,
	\label{eq:fit}
\end{equation}
where the vectors $\mathbf{d}$ and $\mathbf{m}$ have elements $d_n=\mu'_{2,n}/\Delta\mu'_{2,n}$ and $m_n=\mu_{2,n}/\Delta\mu'_{2,n}$, with $\mu'_{2,n}\equiv\sqrt{V_n^2+\sigma_n^2}$ the $N$ measured values and $\Delta\mu'_{2,n}$ the corresponding errors, while $\mu_{2,n}$ are the model predictions for $(M/L)=1$.

\begin{figure}
  \plotone{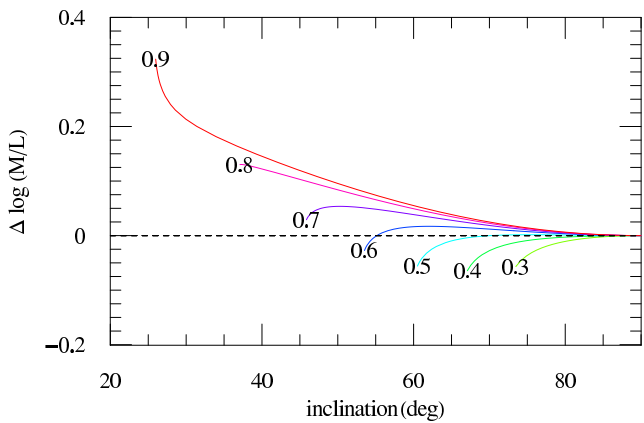}
    \caption{Relative variation of $M/L$ predicted from the virial equations as a function of inclination, for MGE models with different constant observed axial ratios $q'$ going from an observed flat galaxy $q'=0.3$ (green line) to an observed round object $q'=0.9$ (red line). At each observed axial ratio the density is assumed to be rounder than an extreme value of $q=0.1$. Note the small variation $\la10$\% of the $M/L$ estimate at different inclinations, for $q'\la0.7$. }
    \label{fig:ml_variation}
\end{figure}

As previously noticed by other authors \citep[e.g.][]{vdm91} the $M/L$ derived from Jeans modeling is only weakly dependent on inclination, due to the fact that the increased flattening of a model at low inclination ($i=90^\circ$ being edge-on) is compensated by a decrease of the observed velocities, due to projection effects. A quantitative explanation can be obtained by studying how the $M/L$ determined from the tensor virial equations varies with inclination, for given observed surface brightness and $\mu_2$. This variation for an MGE model is given by equation~(77) of \citet{ems94}. If the density is stratified on similar spheroids, the $M/L$ variation is independent of the radial density profile, while only being a function of the observed axial ratio of the galaxy and the assumed inclination. A plot of the variation of $M/L$ with inclination, for different observed axial ratios, is presented in \reffig{fig:ml_variation} and shows that one should expect errors smaller than $\la10$\% even for significant errors in the inclination, if the observed galaxy has an observed axis ratio $q'$ less than 0.7. When the galaxy is rounder than this, it is crucial to obtain a good value for the inclination, otherwise the measured $M/L$ can be significantly in error.

The best-fitting inclination provided by the two-integral models needs not necessarily be correct, as there is no physical reason why the real galaxies should be well described by two-integral models. Nonetheless we found that for the four galaxies in our sample which are not already constrained by the photometry to be nearly edge-on, and for which an independent estimate of the inclination is provided by the geometry of the dust or by the kinematics of the gas (NGC~524, NGC~2974, NGC~4150 and NGC~4459), the inclination derived from the Jeans model agrees with that indicated by the dust or gas (see Appendix~A). Moreover, the round non-rotating giant elliptical galaxies (e.g. NGC~4552 and NGC~4486) are best fitted for an inclination of $90^\circ$, indicating that they are intrinsically close to spherical. This is in agreement with the fact that this whole class of massive galaxies, with flat nuclear surface-brightness profiles, which are often found at the centre of galaxy clusters, always appear nearly round on the sky and never as flattened and fast-rotating systems. They cannot all be flat systems seen nearly face-on, as the observed fraction is too high \citep[e.g.][]{tre96}.

\subsection{Influence of a dark-matter halo}
\label{sec:dark-matter}

In the previous Section we derived the mass density for the dynamical models by deprojecting the observed stellar photometry under the simplifying assumption of a constant $M/L$. In the case of spiral galaxies, for which tracers of the potential are easy to observe, a number of observational and theoretical lines of evidence suggest that, either they also contain a substantial dark matter component \citep[see][for a recent review]{bin04}, or that a modification to the Newtonian gravity law is needed \citep{mil83,bek04}. The situation is less clear regarding early-type galaxies, where it appears that in at least some galaxies dark matter is important \citep{car95,rix97,ger01}, while in others it may not be present at all \citep{rom03,fer05}.  In the simple case in which the dark matter density distribution follows the stellar density, the adopted dynamical models will be correct, but the measured $M/L$ will include the contribution of the stellar component as well as the one of the dark matter. However in general the dark matter profile may not follow the stellar density, so the assumptions of our models could be fundamentally incorrect.

We briefly investigate the effect of a possible dark matter halo on our results using a simple galaxy model. We assume the galaxy to be spherically symmetric and the stellar density to be described by a \citet{her90} profile (of total mass $M=1$), which reasonably well approximates the density of real early-type galaxies:
\begin{equation}
	\rho_\star(r)=\frac{1}{2\pi}\frac{1}{r(r+1)^3}.
\end{equation}
Here the density goes as $\rho_\star\propto r^{-1}$ for $r\ll1$ and as $\rho_\star\propto r^{-4}$ for $r\gg1$.  The dark matter contribution is represented by a logarithmic potential,
\begin{equation}
	\Phi_{\rm DM}(r)=\frac{v_0^2}{2} \ln(r^2+r_0^2),
\end{equation}
which is the simplest potential producing a flat circular velocity $v_0$ at large radii ($r\gg r_0$), as observed in real galaxies.

With these assumptions, the projected second moment $\mu_2$ of an isotropic model, which is equal to the projected velocity dispersion $\sigma_p$ due to the spherical symmetry, is given, e.g., by equation (29) of \citet{tre94}. Substituting the adopted expressions, we obtain:
\begin{eqnarray}
\lefteqn{\sigma_p^2(R)=\frac{1}{\pi\Upsilon I(R)}\times}\nonumber\\
& & 	\int_{R}^{\infty} \left[\frac{1}{r(r+1)^2}
        +\frac{v_0^2}{(r^2+r_0^2)}\right]\frac{\sqrt{r^2-R^2}}{(r+1)^3}
        \; \mathrm{d} r.
\end{eqnarray}
where $I(R)$ is the stellar surface brightness at the projected radius $R$, $\Upsilon$ is the stellar $M/L$, and we assumed a gravitational constant $G=1$. This integral can be evaluated analytically \citep[$I(R)$ is given in][]{her90}, but it is trivially computed numerically.

\begin{figure}
  \plotone{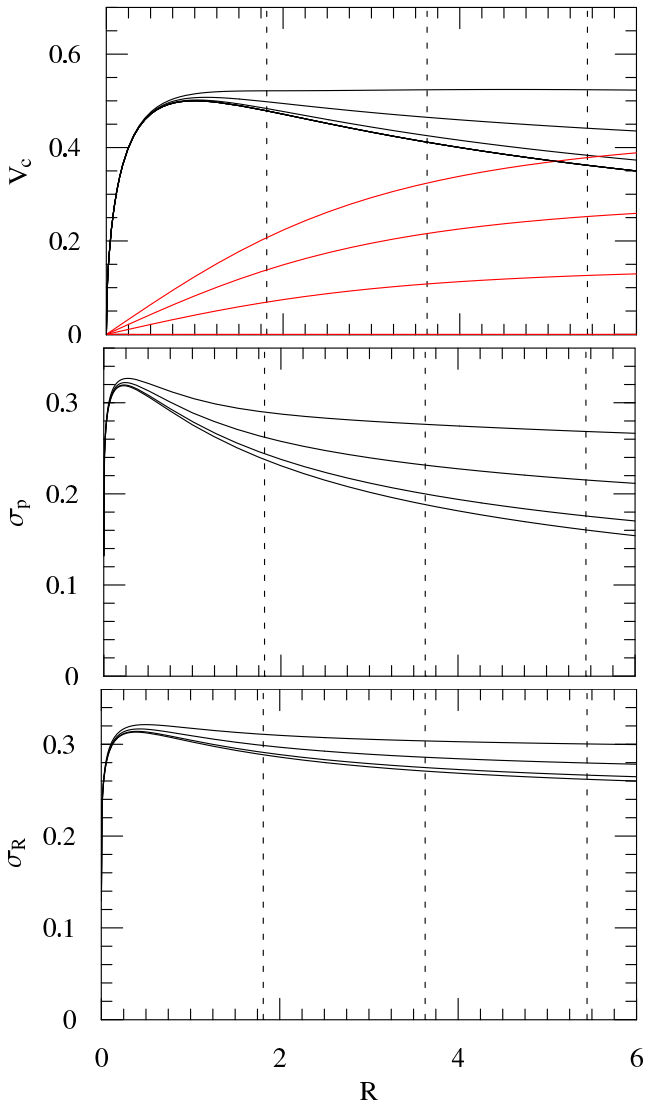}
    \caption{{\em Top Panel:} black lines, from bottom to top, circular velocity $v_c$, as a function of radius $R$, of a spherical Hernquist galaxy model, with a dark halo described by a logarithmic potential, for four different values of the asymptotic halo velocity $v_0=0,0.15,0.30,0.45$. The contribution to the circular velocity of the dark halo component alone is shown with the red lines. The three vertical dashed lines indicate the position of 1, 2 and 3 half-light radii of the model. {\em Middle Panel:} Projected velocity dispersion $\sigma_p$ for the same four models shown in the top panel. The lowest curve is the Hernquist model without a dark halo ($v_0=0$). {\em Bottom Panel:} same as in the middle panel, for the luminosity-weighted aperture velocity dispersion $\sigma_{\rm R}$. This is the same quantity which is plotted in \reffig{fig:sigma_aper} for the real galaxies.}
    \label{fig:dark_halo}
\end{figure}

The results from this formula, evaluated for four different values of $v_0=0,\ldots,v_{0,{\rm max}}$, are shown in the bottom panel of \reffig{fig:dark_halo}. The characteristic radius $r_0=3.5$ of the dark halo and the value of $v_{0,{\rm max}}=0.45$ were chosen so as to make the circular velocity $v_c$ of the stellar plus dark matter component as constant as possible in the range $R=1.815-5.446$ (top panel of \reffig{fig:dark_halo}), which corresponds to 1--3 half-light radii $R_{\rm e}$. This simple choice tries to mimic the typical observations of real early-type galaxies \citep[e.g.\ Figure~2 of][]{ger01}. Under these assumptions the dark matter represents 16\% of the total mass inside $R_{\rm e}$ and 52\% inside $3 R_{\rm e}$. These values are roughly consistent with the values reported by \citet{ger01} for real galaxies.

For this simple model, and assuming the largest dark matter contribution, the projected velocity dispersion $\sigma_p$ is increased by 22\% at $R_{\rm e}$ (the typical maximum radius sampled by our \sauron\ kinematics) and by 9\% at $R_{\rm e}/2$, with respect to a model without dark halo. The luminosity-weighted velocity dispersion inside a circular aperture $\sigma_{\rm R}$ is increased by 8\% inside $R_{\rm e}$ and by 4\% inside $R_{\rm e}/2$ (bottom panel of \reffig{fig:dark_halo}).

To get an indication of the overestimation of the stellar $M/L$ we would obtain by fitting constant $M/L$ models to galaxies with dark halos, we logarithmically rebinned the $\sigma_p$ profile of the model with dark halo (to mimic the spatial sampling of the real data), and fitted to it the model without dark halo, in the radial range 0--$R_{\rm e}$. We obtained that the true stellar $M/L$ would be overestimated by $\sim12\%$, in the case of maximum dark halo. This shows that, under these hypotheses, the dark matter would start producing small but measurable variations in the global $M/L$.

\subsection{Three-integral dynamical modeling}
\label{sec:schwarzschild}

As we mentioned in the introduction, part of the observed FP tilt could come from non-homology in the orbital distribution of galaxies, as a function of luminosity. To test if this is the case, we compared the $M/L$ derived using the quicker and simpler two-integral Jeans models, with general stationary and axisymmetric three-integral dynamical models.

The three-integral dynamical modeling we adopted is based on Schwarzschild's (1979) numerical orbit-superposition method, which is able to fit all kinematic and photometric observations \citep{ric88,rix97,vdm98}. A similar approach was adopted by us and by other groups to measure the black hole (BH) masses in galaxy nuclei or to analyse the stellar orbital distribution \citep[e.g.][]{cre99,cap02a,ver02,geb03,val04,thoj04}. The method provides a general description of axisymmetric galaxies, its stronger assumption being that of the stationarity of the galaxy potential. In practice the method only requires the potential not to vary on the time-scale required to sample the density distribution of an orbit. As the dynamical time-scale in ellipticals is of the order of a few $10^7$ yr at $R_{\rm e}$, while galaxy evolution generally happens on much longer time-scales, this assumption is expected to be valid in the regions that we sample with our kinematics.

\begin{figure}
  \plotone{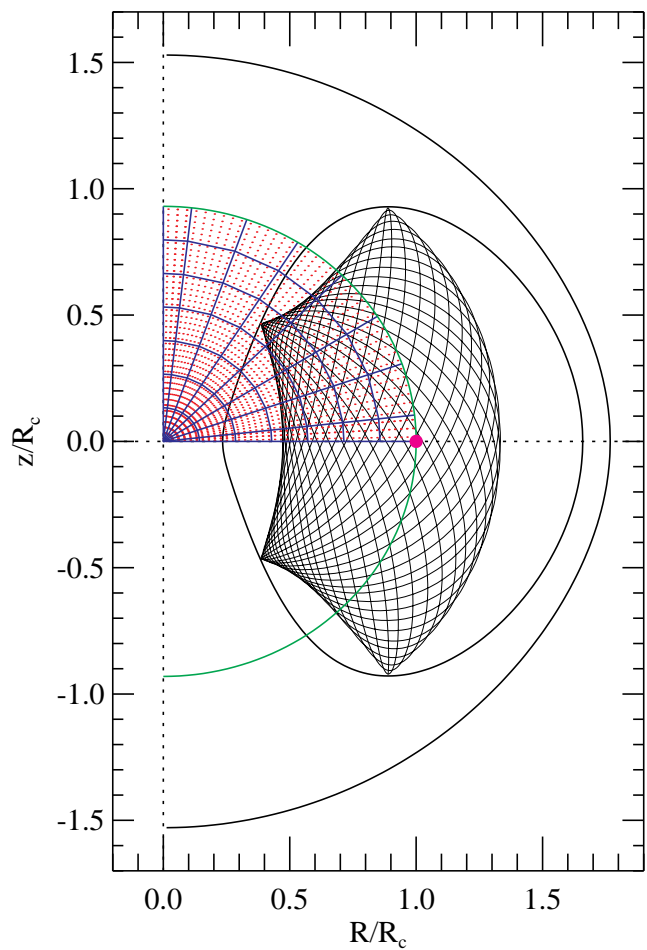}
    \caption{Visualization of our orbital starting scheme and dithering at a given energy $E$ in the meridional plane $(R,z)$ of an axisymmetric galaxy \citep[cf. Fig.~5 of][]{vdm98}. The large magenta dot marks the location of the circular orbit.  At each $E$ we start orbits from the meridional plane with $v_R=v_z=0$ and $L_z>0$, with initial coordinates arranged in a polar grid, linear in the eccentric anomaly and in radius (small red dots), going from $R=z=0$, to the curve defined by the locus (green line) of the thin tube orbits, to reduce duplications. We then construct `dithered' orbital bundles by grouping together (blue lines) every adjacent set of $N_{\rm dither}^3$ orbits and coadding them before computing the observables (at one $E$ one can only visualize $N_{\rm dither}^2$ points or 6$\times$6 in this plot). The first and last angular sectors have half size, to provide a uniform angular sampling across the $R$ and $z$ axes (considering the bi-symmetry of the model). Also shown is a representative regular non-resonant orbit, started near the centre of one orbital bundle, enclosed in its zero-velocity curve. The outermost black solid line is the equipotential, which includes every orbit at the given $E$.}
    \label{starting_scheme}
\end{figure}

In our implementation of the Schwarzschild method we use the MGE parameterization to describe the stellar density and to calculate the gravitational potential, as in \citet{cap02a} and \citet{ver02}. However, we use here a new code which was specifically designed with the improved quality of our input integral-field kinematical data in mind. We verified that the new code produces the same results as the old one (within the numerical approximations of the method), when identical conditions are adopted. The key updates to the modeling method, with respect to the description given in \citet{vdm98} and in \citet{cap02a}, consist of:
\begin{enumerate}
	\item The orbit library required by the method has to be sampled by choosing the starting points of the orbits so as to cover a three dimensional space defined by the isolating integrals of motion, the energy E, the $z$-component of the angular momentum $L_z$ and the third integral $I_3$. In previous works, at each $E$, we sampled orbits linearly in $L_z$, and for each $(E,L_z)$ pair, an angle was used to parameterise $I_3$ \citep[see Fig.~5 of][]{vdm98}. Here at each $E$ we start orbits from the meridional plane with $v_R=v_z=0$ and $L_z>0$ with positive initial coordinates arranged in a polar grid (linear in eccentric anomaly and in radius) going from $R=z=0$, to the curve defined by the locus of the thin tube orbits (\reffig{starting_scheme}), to reduce duplications. The reason for our change is that we want to sample observable space as uniformly as possible, as opposed to sample the space of the integrals of motion, which is not directly observable. With the new scheme the cusp singularities, which characterize the observables of orbits \citep[see Fig.~1.2 of][]{cap04} are distributed evenly on the sky plane.

	\item Ideally the orbit library in Schwarzschild's method should provide a complete set of basis functions for the DF. Individual orbits corresponds to DFs that are $\delta$-functions and an infinite number of them can indeed represent any smooth DF. However only a limited number of orbits can be used in real models, which then have a non-smooth DF and generate unrealistically non-smooth observables. To alleviate this problem, instead of using single orbits as basis functions for the DF in the models, we use `dithered' orbital components, constructed by coadding a large bundle of orbits started from different but adjacent initial conditions. The starting points are arranged in such a way that the (regular) orbits differ in all three integrals of motion (\reffig{starting_scheme}). This technique produces smoother components, which are also more representative of the space of integrals of motions they are meant to describe \citep[a similar approach was used for spherical models by][]{ric88,rix97}. With dithered orbits we can construct models which better approximate real galaxies than would otherwise be possible with single orbits, due to computational limitation. Dithering is crucial for this work, as otherwise the numerical noise generated by the relatively small orbit number would limit the accuracy in the determination of the $M/L$.

	\item We do \emph{not} use Fourier convolutions to model PSF effects. Given that the computation of the orbital observables is intrinsically a Monte Carlo integration, we also deal with the PSF in the same way. The phase space coordinates are computed at equal time steps during the orbit integration \citep[using the DOP853 routine by][]{hai93}. They are projected on the sky plane, generating the triplets of values $(x',y',v_z')$, which are regarded as `photons' coming from the galaxy. The sky coordinates $(x',y')$ of each photon are randomly perturbed, with probability described by the MGE PSF, before being recorded directly into any of the observational apertures. In this way no interpolation or intermediate grids are required and the accuracy of the computation of the observables is only determined by the (large) number of photons generated.  Voronoi 2D-binning is treated in the model in the same way as in the observed data (\refsec{sec:kinematics}).
\end{enumerate}

For the sampling of the orbit library in the present paper we adopted a grid of 21 energies, and at every energy we used 8 angular and 7 radial sectors, with $N_{\rm dither}=6$ (as in \reffig{starting_scheme}). This means that the final galaxy model is made of $21\times(8-1)\times7\times2\times6^3=444,528$ different orbits, which are bundled in groups of $6^3=216$ before the linear orbital superposition. From each orbital component we generate $\sim2\times10^{8}$ `photons' on the sky plane, which are used for an accurate Monte Carlo calculation of the observables inside the Voronoi bins. At each bin position we fit $V$, $\sigma$ and four Gauss-Hermite parameters up to $h_6$.

Some of the models whose $M/L$ we use in this paper were already presented elsewhere. An application of this new code to the modeling of the 2D-binned kinematics of the elliptical galaxy NGC~4473, which contains two counterrotating spheroids, and a study of the orbital distribution of the giant elliptical galaxy M87 was presented in \citet{cap05}. The modeling of the elliptical galaxy NGC~2974, and detailed tests of the ability of this orbital sampling scheme to recover a given $M/L$ and a realistic DF are presented in \citet{kra05}.

In contrast to studies of spiral galaxies, where the inclination can be inferred from the observed geometry of the disk, there is usually no obvious way to infer the inclination of early-type galaxies, and it therefore is one of the free parameters of the models. \citet{kra05} showed that there is evidence for the inclination to be possibly degenerate using axisymmetric three-integral models, even if the LOSVD is known at every position on the projected image of the galaxy on the sky. Here we confirmed this result using a larger sample of galaxies, in the sense that the models are generally able to fit the observations within the measurement errors, for wide ranges of variation in inclination. For these reasons we fitted the $M/L$ by constructing sequences of models with varying $M/L$, while keeping the inclination fixed to the best-fitting value provided by the Jeans models (Table~\ref{tab1}). We show in Appendix~A that the inclination derived from the Jeans models may be more accurate than just assuming all galaxies are edge-on, although we are aware it may not always be correct. In any case, we have shown in \reffig{fig:ml_variation} that the results of this work do not depend critically on the assumed inclination.

The Schwarzschild models use the same MGE parameters as the Jeans models (\refsec{sec:jeans}) and also include a central BH as predicted by the $M_{\rm BH}$--$\sigma$ relation. More details on the modeling and the motivation of the changes, together with a full analysis of the orbital distribution inferred from the models are given in a follow-up paper. The fact that the models can generally fit the observed density and the \sauron\ integral-field kinematics of our galaxies with a constant $M/L$ seems to confirm the suggestion of \refsec{sec:dark-matter}: either mass follows light in galaxy centres, or we cannot place useful constraints on the dark-matter halo from these data. This may be explained by the existence of an intrinsic degeneracy in the simultaneous recovery of the potential and the orbital distribution from the observed LOSVD, as discussed e.g.\ in Section~3 of \citet{val04}, combined with the relatively limited spatial coverage. In a few galaxies the $h_4$ and $h_6$ Gauss-Hermite moments cannot be reproduced in detail by our models and may be improved e.g.\ by allowing the $M/L$ to vary or extending to triaxial geometry.

\section{Results}
\label{results}

\subsection{Comparing $M/L$ of two and three-integral models}
\label{sec:2i-3i}

\begin{figure}
  \centering\includegraphics[width=0.75\columnwidth]{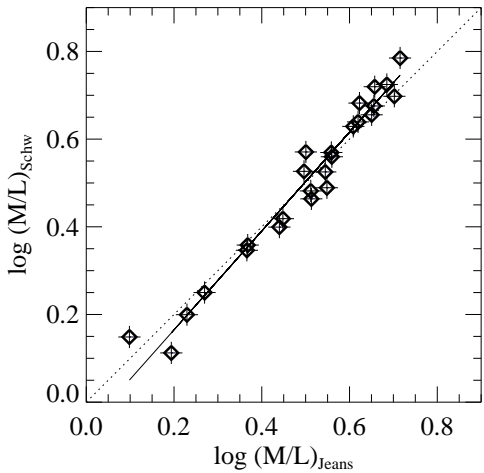}
    \caption{Comparison between the 25 $M/L$ determinations using the two-integral Jeans dynamical models and the three-integral Schwarzschild models. An error of 6\% in the model accuracy is required to explain the observed scatter along the best-fitting relation (solid line). The dotted line represents a one-to-one relation.}
    \label{fig:ml-jeans-schwarzschild}
\end{figure}

In this Section we compare the measurements of $M/L$ obtained with the two-integral Jeans models described in \refsec{sec:jeans} with the results obtained with the more general Schwarzschild models of \refsec{sec:schwarzschild}. The usefulness of this comparison comes from the fact that the Jeans models are not general, but can be computed to nearly machine numerical accuracy, so they will generally provide biased results, but with negligible numerical noise. The Schwarzschild models are general, within the assumptions of stationarity and axisymmetry, but are affected by numerical noise and can be influenced by a number of implementation details. A comparison between these two substantially different modeling methods will provide a robust estimate of the uncertainties in the derived $M/L$, including systematic effects. The best-fitting correlation (\reffig{fig:ml-jeans-schwarzschild}) shows a small systematic trend $(M/L)_{\rm Schw}\propto (M/L)_{\rm Jeans}^{1.13\pm0.05}$, while the scatter in the relation indicates an intrinsic error of 6\% in the models. Formal errors in the Schwarzschild $M/L$ are generally not too different from this more robust estimate \citep[e.g.][]{kra05}.

One can see in \reffig{fig:ml-jeans-schwarzschild} that the galaxies with the highest $M/L$ tend to show an $(M/L)_{\rm Schw}$ which is systematically higher than $(M/L)_{\rm Jeans}$ (see also Table~\ref{tab1}). The difference in the $M/L$ is likely due mainly to the fact that the Schwarzschild models use the full information on the LOSVD, while the Jeans models are restricted to the first two moments. For this reason, in the following Sections, unless otherwise indicated, we will use the term ``dynamical $M/L$'', or simply $M/L$, to refer to the $(M/L)_{\rm Schw}$, which is expected to provide a less biased estimate of the true $M/L$.

\subsection{Observed correlations}
\label{sec:corr}

The data consists of $N=25$ estimates of the dynamical $M/L$ in the $I$-band and corresponding measurements of the luminosity-weighted second moment of the LOSVD $\sigma_{\rm e}$ within the half-light radius.  We fitted a linear relation of the form $y=\alpha+\beta x$, with $x=\log(\sigma/\sigma_0)$ and $y=\log(M/L)$, and minimizing
\begin{equation}
	\chi^2=\sum_{j=1}^N \frac{(y_j-\alpha-\beta x_j)^2}
	{(\Delta y_j)^2 + \beta^2 (\Delta x_j)^2},
\end{equation}
with symmetric errors $\Delta x_j$ and $\Delta y_j$, using the FITEXY routine taken from the IDL Astro-Library \citep{lan93}, which is based on a similar routine by \citet{pre92}. This is the same approach adopted by \citet{tre02} to measure the slope of the correlation between the mass $M_{\rm BH}$ of the central BH and $\sigma_{\rm e}$, and we refer to that paper for a comparison of the relative merits of alternative methods.  The $M/L$ measurements are inversely proportional to the assumed distance, which is taken from \citet{ton01} and was adjusted to the new Cepheids zeropoint of \citet{fre01} by {\em subtracting} 0.06 mag from the SBF distance moduli (Table~\ref{tab1}). For this reason the errors in the distance represent a lower limit to the errors in $M/L$. In the fit, we assigned to the $M/L$ the distance errors quadratically coadded to the 6\% modeling error derived in \refsec{sec:2i-3i}. The $\sigma_{\rm e}$ values are measured from spectra of extremely high $S/N$, and have negligible statistical errors, but their accuracy is limited by template mismatch and possible calibration accuracy. We adopted an error in $\sigma_{\rm e}$ of 5\%. An upper limit to the error in $\sigma_{\rm e}$ can be derived by comparing our values with the central $\sigma_{\rm c}$ in published data (HyperLeda;\footnote{http://leda.univ-lyon1.fr/} \citealt{pru96}), which is closely related to our $\sigma_{\rm e}$, but it is measured in a very different way. We found that errors of 7\% in $\sigma$ are needed to explain the scatter with respect to a linear relation between $\log\sigma_{\rm c}-\log\sigma_{\rm e}$. This scatter also includes very significant systematics due to the differences in the data, analysis and measurement method. The adopted 5\% error provides a conservative estimate of the true errors. Note that for the galaxy M87 our $\sigma_{\rm e}$ value is 20\% smaller than the value adopted by \citet{tre02}. Our \sauron\ kinematics however agrees well with both the G-band and the Ca triplet stellar kinematics by \citet{vdm94}. Adopting our smaller $\sigma_{\rm e}$ M87 would become an outlier of the $M_{\rm BH}-\sigma_{\rm e}$ correlation.

\begin{figure}
  \plotone{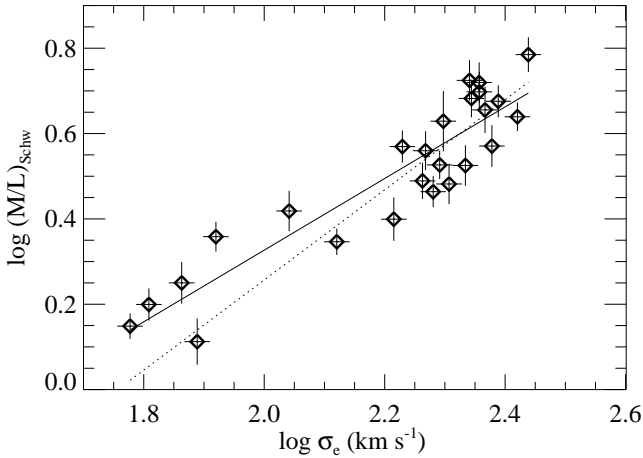}
    \caption{$(M/L)$--$\sigma_{\rm e}$ correlation. The errors in the $M/L$ values shown here and in the following plots represent the uncertainty in the distance as given by \citet{ton01}, quadratically coadded to the 6\% modeling error determined in \refsec{sec:2i-3i}.  The errors on $\sigma_{\rm e}$ are assumed to be 5\%. The solid line is the correlation obtained by fitting all galaxies, while the dotted line is the fit obtained by excluding the five galaxies with $\log\sigma_{\rm e}<2$. The galaxy with the smallest $\sigma_{\rm e}$ is M32.}
    \label{fig:ml-sigma}
\end{figure}

We find a tight correlation (\reffig{fig:ml-sigma}) with an {\em observed} rms scatter in $M/L$ of 18\%. The best-fitting relation is
\begin{equation}
(M/L) = (3.80\pm0.14) \left(\frac{\sigma_{\rm e}}{200\; \kms}
                                                \right)^{0.84\pm0.07}.
\label{eq:ml-sigma}
\end{equation}
The value of $\sigma_0=200\;\kms$ was chosen, following \citet{tre02}, to minimize the uncertainty in $\alpha$ and the correlation between $\alpha$ and $\beta$. The best-fitting values and the uncertainties, here as in all the following correlations, were determined after increasing the errors $\Delta y_j$ in the $M/L$ by quadratically coadding a constant `intrinsic' scatter so that $(\Delta y_j)^2$ is replaced by $(\Delta y_j)^2+(\Delta y_0)^2$, to make $\chi^2/\nu=1$, where $\nu=N-2=23$ is the number of degrees of freedom in the linear fit \citep[see][for a discussion of the approach]{tre02}. The constant intrinsic scatter implied by the observed scatter in this correlation is $\sim13\%$ ($\Delta y_0=0.053$).

To test the robustness of the slope we also performed a fit only to the galaxies with $\log\sigma_{\rm e}>2$, which is the range most densely sampled by FP studies \citep[e.g.][]{jor96}. The relation becomes in this case $(M/L)\propto \sigma_{\rm e}^{1.06\pm0.18}$, which is steeper, but is still consistent with equation~(\ref{eq:ml-sigma}) within the much larger uncertainty. The steepening of the correlation at high $\sigma_{\rm e}$ may also be due to a non perfect linearity of the relation \citep[e.g.][]{zar05}, but this cannot be tested with our limited number of galaxies. The correlation computed using the Jeans determinations is even tighter than equation~(\ref{eq:ml-sigma}), and gives $(M/L)_{\rm Jeans}\propto \sigma_{\rm e}^{0.77\pm0.06}$ with an observed scatter of only 15\%. However, due to the possible bias of the Jeans $M/L$ values, we will use the Schwarzschild determinations in what follows.

The use of $\sigma_{\rm e}$ determined from integral-field data and using a large aperture has the significant advantage that it is conceptually rigorous and accurate. However the $(M/L)$--$\sigma$ correlation does not appear to depend very strongly on the details by which $\sigma$ is determined. We performed the same fit of \reffig{fig:ml-sigma} using as value of the dispersion the $\sigma_{\rm e/8}$ measured from the \sauron\ data in an aperture of radius $R_{\rm e}/8$. In this case the best-fitting correlation has the form $(M/L)\propto \sigma_{\rm e/8}^{0.75\pm0.06}$ and the observed rms scatter in $M/L$ is 21\%. As an extreme test we used as dispersion the inhomogeneous set of central values $\sigma_{\rm c}$, obtained from long-slit spectroscopic observations, as given in the HyperLeda catalogue. In this case the best-fitting correlation becomes $(M/L)\propto \sigma_{\rm c}^{0.87\pm0.07}$ and has an observed scatter of 20\%. Both correlations are consistent within the errors with the best fitting $(M/L)$--$\sigma_{\rm e}$ correlation, although they have a larger scatter.

We also determined the correlation between the $M/L$ and the second moment of the velocity, corrected with equation~(18) of \citet{vdm93}
\begin{equation}
\tilde{\sigma}_{\rm e}\approx\sigma_{\rm e}(1+h_{\rm 4,e}\sqrt{6}),
\label{eq:corr_h4}
\end{equation}
to include the contribution to the second moment of a nonzero $h_{\rm 4,e}$ parameter. Equation~(\ref{eq:corr_h4}) was obtained by integrating the LOSVD to infinite velocities. In practice the second moment of a parameterised LOSVD with dispersion $\sigma$ and $h_4=0.1$ is equal to $\tilde{\sigma}=(1, 1.11, 1.21)\times\sigma$ if one sets the LOSVD to zero for velocities larger than $|V|>(2.45, 3, 4)\times\sigma$ respectively. This shows that the correction has to be used with care, being highly sensitive to the details of the LOSVD at large velocities, where the LOSVD cannot be determined reliably. In fact a measured positive $h_4$ parameter, which means the LOSVD is narrower than a Gaussian at small velocities, does not necessarily imply that the wings of the LOSVD are precisely as described by the Gauss-Hermite parameterization. We find a correlation consistent, within the errors, with equation~(\ref{eq:ml-sigma}) and with comparable scatter.

\begin{figure}
  \plotone{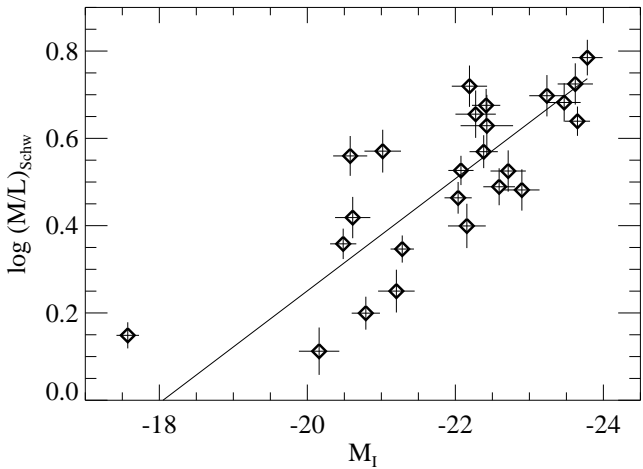}
    \caption{$(M/L)$--$L_I$ correlation. Note that all galaxies, with the exception of M32 are brighter than the limit $M_I\approx-20$ mag, in absolute magnitudes, which is set by the luminosity selection criterion of the \sauron\ survey. Due to the sensitivity of the fit to the single $M/L$ of M32, this galaxy was not included in this and in the following fits involving luminosity or mass, but it is included in the plots, to allow a comparison with \reffig{fig:ml-sigma}.}
    \label{fig:ml-lum}
\end{figure}

We also studied the correlation of the dynamical $M/L$ with the galaxy luminosity (\reffig{fig:ml-lum}). For this we used our own total magnitudes in the $I$-band $I_T$ given in Table~\ref{tab1}, transformed into absolute magnitudes using the SBF distances and the galactic absorption \citep{sch98} given in Table~\ref{tab1}. The $I_T$ measurements are dominated by possible systematic errors due to the extrapolation involved and the possible contamination by nearby objects. To obtain a realistic estimate of the errors we compared our $I_T$ with the same value\footnote{$I_T=\bar{m}_I-\bar{N}_I$ in the notations of Table~1 of \citet{ton01}} measured by \citet{ton01} in the same photometric band. A fit to the two sets of values showed no systematic trend with magnitude and a zeropoint shift of $0.09$ mag. The observed scatter is consistent with errors of 13\% in each measurement, and we assigned this error to our $I_T$ values. A stringent internal test on the errors in the $I_T$ measurements comes from the comparisons with the total luminosities obtained by coadding all the Gaussians of the MGE models of \refsec{sec:mge}. A correlation of the two quantities shows no deviant points and no systematic trend in either slope or zeropoint, and is fully explained by an 8\% random error in both quantities. This estimate is similar to the typical measurement errors of 9\% reported by \citet{jor96}. We also compared $I_T$ with the corresponding values of $B_T$ in the $B$-band as given by the HyperLeda catalogue \citep{pru96}. We found a smooth variation of the $B-I$ colour with luminosity, consistent with the quoted measurements errors in this quantity and indicating the absence of significant systematic errors or any deviant point.  The fit was performed in the same way as above and provided the relation:
\begin{equation}
(M/L) = (2.35\pm0.19) \left(\frac{L_I}{10^{10} L_{I,\sun}}\right)^{0.32\pm0.06}
\label{eq:ml-lum}
\end{equation}
This correlation has an observed rms scatter in $M/L$ of 31\% which is much larger than that of the correlation (\ref{eq:ml-sigma}). In particular there is a significant scatter for the eight galaxies in the luminosity range $-21.5 \la M_I\la -19.5$ mag.  Given the fact that all galaxies in the \sauron\ sample were selected to have $M_B\le-18$ mag, which translates in the $I$-band into $M_I\la-20$ mag, the fit depends very sensitively on the single $M/L$ of M32, which also has a small distance error. Moreover there is some indication that the dwarf E galaxies like M32 may populate a different FP from the normal E galaxies (\citealt{bur97}; but see \citealt{gra03}). For these reason we excluded M32 from this fit and all the following fits involving luminosity or mass.

\subsection{Understanding the tightness of the correlations}

The decrease of the scatter in $M/L$, when going from the $(M/L)$--$L$ correlation to the $(M/L)$--$\sigma$ one is reminiscent of what happens to the similar correlations involving the BH mass $M_{\rm BH}$ instead of the $M/L$. This decrease of the scatter shows that at a given galaxy luminosity the $M/L$ is higher when the $\sigma$ is higher. One may apply the arguments used to understand the tightness of the $M_{\rm BH}$--$\sigma$ relation \citep{geb00,fer00} to the $(M/L)$--$\sigma$ relation as well. There can be two main reasons for the $(M/L)$--$\sigma$ relation to be tighter than the $(M/L)$--$L$ relation: (i) the $M/L$ may correlate with galaxy mass, but the luminosity may not be a good mass estimator, due to differences of $M/L$ at a given luminosity; in this case variations of $\sigma$ would reflect variations of $M/L$; (ii) the $M/L$ may be related to the galaxy compactness, which implies a higher $\sigma$ at a given mass. In the case of the $M_{\rm BH}$--$\sigma$ relation both effects seem to play a role: \citet{kor01} showed that the galaxies with higher $\sigma$ are indeed more compact than average \citep[see also][]{gra01}. Subsequently \citet{mar03} and \citet{har04} showed that much tighter correlations for $M_{\rm BH}$ are obtained using the galaxy mass instead of the luminosity.

\begin{figure}
  \plotone{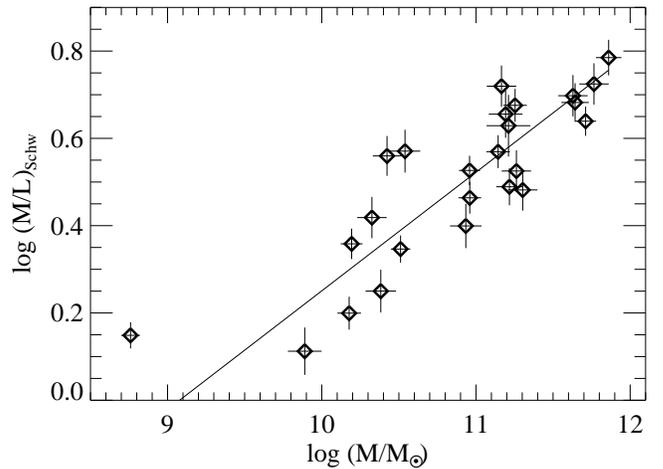}
    \caption{$(M/L)$--$M$ correlation.  Here the $I$-band luminosity of each galaxy was converted into mass by multiplication by its $M/L$. The least massive galaxy is M32, and was not included in the fit.}
    \label{fig:ml-mass}
\end{figure}

It is clear from \reffig{fig:ml-lum} that there is a spread in $M/L$ at a given galaxy luminosity, so the first argument above must play a role in tightening the correlation. To test the importance of this effect we verify how much better the $M/L$ correlates with galaxy mass after multiplying each luminosity by its measured $M/L$. The best-fitting correlation (\reffig{fig:ml-mass}) is given by
\begin{equation}
(M/L) = (1.78\pm0.16) \left(\frac{M}{10^{10} M_{\sun}}\right)^{0.27\pm0.03}.
\label{eq:ml-mass}
\end{equation}
This correlation has an observed scatter in $M/L$ of 24\%. It is tighter than the $(M/L)$--$L$ correlation, but still significantly less tight than the $(M/L)$--$\sigma$ one.

An alternative way to verify the importance of the $M/L$ variations in increasing the scatter of the $(M/L)$--$L$ relation is to use the galaxy luminosity in the near infrared, instead of in the $I$-band. The near infrared better samples the light emitted by the bulk of the galaxies' old stellar population. Moreover in that wavelength range the $M/L$ variations with mass are smaller (e.g.\ \citealt{bel01}). These are apparently the reasons why \citet{mar03} found a much tighter $M_{\rm BH}$--$L$ correlation in the $K$-band than in the $B$-band.

\begin{figure}
  \plotone{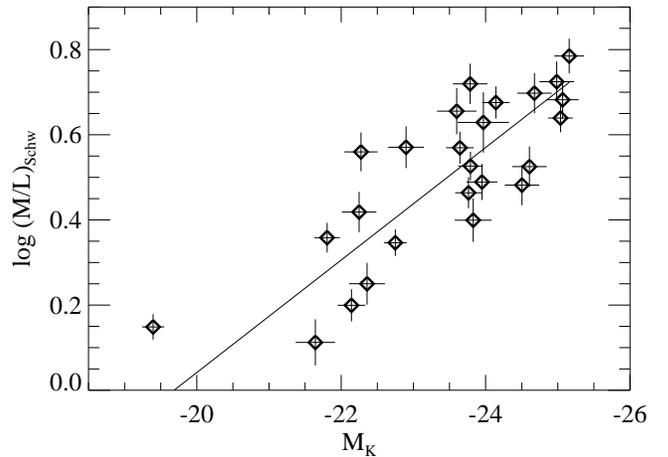}
    \caption{$(M/L)$--$L_K$ correlation. Same as in \reffig{fig:ml-lum}, but here the $M/L$ is plotted against the 2MASS galaxy absolute magnitude in the $K$-band. The faintest galaxy is M32 which was not included in the fit.}
    \label{fig:ml-lum_k}
\end{figure}

We want to see if the same tightening happens in our case with the $M/L$. The correlation between the $I$-band $M/L$ and the $K$-band luminosity, extracted from the 2MASS Extended Source Catalog (XSC; \citealt{jar00}), is shown in \reffig{fig:ml-lum_k}. We adopted the same constant error of 13\% in the luminosity, as we used in the $I$-band, instead of the errors given in the XSC, which have an unrealistically small median value of 2\%. The best fitting relation is nearly identical to equation~(\ref{eq:ml-lum}) having the form
\begin{equation}
(M/L) = (1.88\pm0.20) \left(\frac{L_K}{10^{10}
                            L_{K,\sun}}\right)^{0.32\pm0.05},
\label{eq:ml-lum-k}
\end{equation}
adopting $M_{K,\sun}=3.28$ mag. This relation has indeed a slightly smaller rms scatter in $M/L$, of 28\%, than the $I$-band correlation.

\begin{figure}
  \plotone{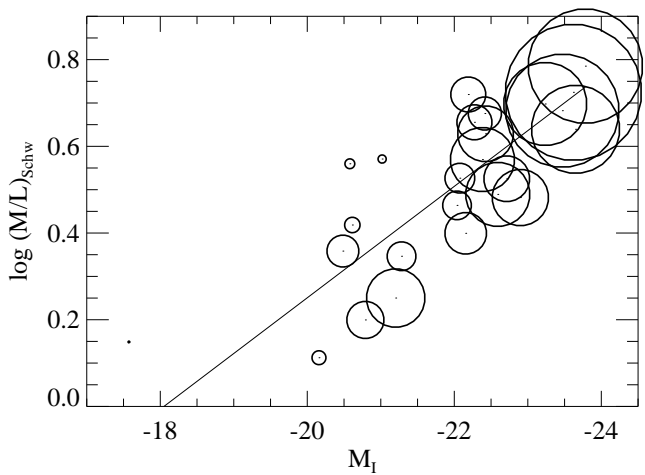}
    \caption{$(M/L)$--$L_I$ correlation as in \reffig{fig:ml-lum}. The radius of the circles, centered around each measurement, is proportional to the galaxy $R_{\rm e}$ in pc.}
    \label{fig:ml-lum-re}
\end{figure}

Finally, to test the importance of the second argument above, namely the effect of the galaxy compactness in tightening our $(M/L)$--$\sigma$ correlation, we show in \reffig{fig:ml-lum-re} the same $(M/L)$--$L$ relation as in \reffig{fig:ml-lum}, but we also overplot the effective radii $R_{\rm e}$ of the galaxies (in pc). The eight galaxies with $-21.5\la M_I\la -19.5$ differ substantially in their $R_{\rm e}$. Indeed the galaxies with higher $M/L$ are also the most compact at a given luminosity. This explains why they also have the highest $\sigma$ (see \citealt{kor01}, for a discussion), and it shows why, at least for our limited galaxy sample, the $(M/L)$--$\sigma$ correlation is even tighter than the $(M/L)$--$M$ relation.

To summarize the results of this Section, we found that (i) the tightest correlation is the $(M/L)$--$\sigma$ relation, followed by $(M/L)$--$M$ and $(M/L)$--$L_K$. The $(M/L)$--$L_I$  relation is the least tight. This sequence of increasing scatter suggests that the $M/L$ correlates primarily with galaxy mass and, at given mass, with galaxy compactness. However we caution the reader that the size of our sample is too small to draw definitive conclusions. In fact, with the exception of the $(M/L)$--$\sigma$ relation, the difference in scatter between the different correlations is only at a few sigma level. A larger sample is needed to make this result stronger.

\subsection{Understanding the tilt of the FP}
\label{sec:fp}

The existence of the FP of the form $R_{\rm e}\propto\sigma^\alpha I_{\rm e}^\beta$, combined with the homology and virial equilibrium assumption $M\propto\sigma^2 R_{\rm e}$ and the geometric definition $L\propto I_{\rm e} R_{\rm e}^2$, yields a prediction for the $(M/L)_{\rm FP}$ \citep{dre87,fab87}. Here we adopt a FP of the form:
\begin{equation}
R_{\rm e}\propto\sigma^{1.24\pm0.07} I_{\rm e}^{-0.82\pm0.02},
\end{equation}
as determined in the Gunn $r$-band by \citet{jor96} from a sample of 225 early-type galaxies in nearby clusters. We choose these numerical values because the photometric parameters $R_{\rm e}$ and $I_{\rm e}$ for this FP were measured from a homogeneous sample of E and S0 galaxies, in the same way as we did (\refsec{sec:phot}). This FP determination agrees within the errors with other determinations in similar photometric bands \citep[see][and references therein]{col01} and also with the $I$-band determination by \citet{sco97}, so the choice of the FP coefficients is not critical. One important exception is the FP determination by \citet{ber03}, derived from a large sample of galaxies from the Sloan Digital Sky Survey. The parameters of that FP are inconsistent with most previous determinations, and the reason for this discrepancy is currently unclear. Considering also that their photometric parameters were derived by fitting the galaxy profiles and not from a growth curve, as in this and other FP works, we decided not to consider their values here. Taking the numerical values by \citet{jor96}, one can derive the following relations for the $(M/L)_{\rm FP}$, as a function of the combination $R_{\rm e}\sigma$ or equivalently galaxy mass, and galaxy luminosity :
\begin{eqnarray}
(M/L)_{\rm FP} & \propto
               & R_{\rm e}^{-1-1/\beta} \sigma^{2+\alpha/\beta}\nonumber\\
& = & R_{\rm e}^{0.22\pm0.03} \sigma^{0.49\pm0.09}\nonumber\\
& \propto & R_{\rm e}^{-(2+\alpha)/(2\beta)-2} M^{1+\alpha/(2\beta)}\nonumber\\
& = & R_{\rm e}^{-0.02\pm0.06} M^{0.24\pm0.05}, \\
(M/L)_{\rm FP} & \propto
             & I_{\rm e}^{-1/2-(1+2\beta)/\alpha} L^{-1/2+1/\alpha}\nonumber\\
& = & I_{\rm e}^{0.02\pm0.04} L^{0.31\pm0.05}.
\label{eq:ml_fp}
\end{eqnarray}
The scatter in the relations amounts to 23\% between $M/L$ and $R_{\rm e}^{0.22}\sigma^{0.49}$ or mass, and 31\% between $M/L$ and $L$. Within the errors the $(M/L)_{\rm FP}$ can be written as a simple power-law of either luminosity or mass alone. This simple dependence is what is usually known as the `tilt' of the FP.  The $(M/L)_{\rm FP}$ cannot be written explicitly as a simple function of $\sigma$, however \citet{jor96} note that ``the inclusion of $R_{\rm e}$ is not even essential'' as a direct correlation between $(M/L)_{\rm FP}$ and $\sigma$ gives a correlation:
\begin{equation}
	(M/L)_{\rm FP}\propto\sigma^{0.86},
\label{eq:fp_sigma}
\end{equation}
with an observed scatter virtually unchanged of 25\%.

As mentioned before, the correctness of the FP prediction for the $M/L$ depends on the homology assumption, regarding both the surface brightness and the orbital distribution of early-type galaxies \citep[e.g.][]{cio96}. Our dynamical $M/L$ is based on detailed photometric models, so it is independent of the spatial homology assumption. Moreover we include the full LOSVD to estimate the $M/L$ and we do not need to make any extra correction for galaxy rotation. Finally the good agreement between the $M/L$ derived with two-integral Jeans and three-integral Schwarzschild models shows that orbital non-homology does not have an important effect on the $M/L$ (\refsec{sec:2i-3i}). We are thus in the position to compare directly our $M/L$ determinations, which do not depend on any homology assumption, with the homology-dependent $(M/L)_{\rm FP}$.

Comparison between the FP tilt given by equation~(\ref{eq:ml_fp}) and  relation~(\ref{eq:ml-lum}) shows that the dependence of the dynamical $M/L$ on luminosity is the same. Taking the measurement errors into account this indicates that non-homology cannot produce more than $\sim15$\% of the observed tilt. For the more reliable correlations involving $\sigma$, comparison of the virial prediction of equation~(\ref{eq:fp_sigma}) and the modeling result of equation~(\ref{eq:ml-sigma}) indicates that non-homology can account for at most $\sim7$\% of the tilt. Both relations consistently imply that the FP tilt reflects essentially a real variation of the total $M/L$ in the central regions of galaxies, which can be due to variations in the galaxies' stellar population and/or in the dark matter fraction. A similar conclusion was reached by \citet{lan04} from general considerations about the observed scaling relations of early-type galaxies. The comparison of this Section however involves uncertainties due to the fact that the tilt may depend on the sample selection criteria. To assess if this plays an important effect we perform in the next Section a direct comparison of the virial predictions for the $M/L$ derived from our own galaxy sample.

\subsection{Comparison with virial predictions of $M/L$}

An alternative way to test the validity of the virial and homology assumptions and their influence on the FP tilt is to compute the `observed' virial $(M/L)_{\rm vir}\propto R_{\rm e}\sigma_{\rm e}^2/L$ and to compare it directly to the $M/L$ derived from the dynamical models. This has the advantage that it can be performed on our own galaxies and does not involve any choice of FP parameters or selection effects. We fitted the correlations of $(M/L)_{\rm vir}$, in the $I$-band, with $\sigma$ and with luminosity, obtaining:
\begin{eqnarray}
 (M/L)_{\rm vir} & \propto & \sigma_{\rm e}^{0.82\pm0.07},\label{eq:ml_vir1}\\
 (M/L)_{\rm vir} & \propto & L^{0.27\pm0.04}.\label{eq:ml_vir2}
\end{eqnarray}
Equations~(\ref{eq:ml_vir1},\ref{eq:ml_vir2}) have an observed rms scatter of 19\% and 27\%, and are fully consistent with the FP determinations in equations~(\ref{eq:fp_sigma}) and (\ref{eq:ml_fp}) respectively. The scatter in these correlations of virial determinations is comparable to the scatter derived using the full dynamical models and, as in that case, appears dominated by the intrinsic scatter in $M/L$.

\begin{figure}
  \centering\includegraphics[width=0.75\columnwidth]{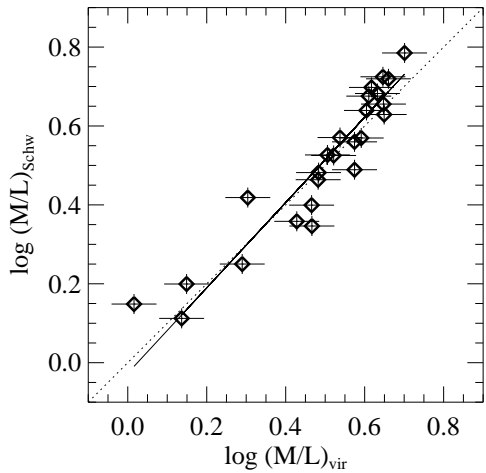}
    \caption{Comparison between the $(M/L)_{\rm vir}=\beta R_{\rm e}\sigma_{\rm e}^2/(L\, G)$ derived from the virial assumption and the $M/L$ obtained from the Schwarzschild models. The values of $(M/L)_{\rm vir}$ were scaled to match the dynamical $M/L$, and the best-fitting factor is $\beta=5.0\pm0.1$. The solid line is a fit between the two quantities, while the dotted line represents a one-to-one correlation.}
    \label{fig:ml-schwarzschild-virial}
\end{figure}

Finally, the most direct way of measuring the accuracy of the homology assumption is to compare $(M/L)_{\rm vir}$ with the $M/L$ from the dynamical models. The correlation is shown in \reffig{fig:ml-schwarzschild-virial} and has the form
\begin{equation}
	(M/L)\propto(M/L)_{\rm vir}^{1.08\pm0.07}.
\label{eq:jeans_vir}
\end{equation}
The observed slope is consistent with the determination based on the FP (\refsec{sec:fp}), implying that both the structural and the orbital non-homology contribution cannot represent more than $\sim15\%$ of the FP tilt (neglecting possible selection effects in our sample, which are very difficult to estimate).  Contrary to all correlations shown before, the scatter in this correlation is not influenced by the errors in the distance, as both $M/L$ estimates use the same distance. Adopting an intrinsic accuracy of 6\% in the $M/L$ determinations (\refsec{sec:2i-3i}), the scatter in $(M/L)_{\rm vir}$ required to make  $\chi^2/\nu=1$ is 14\%. The correlation between the virial and the Jeans estimates gives $(M/L)_{\rm Jeans}\propto(M/L)_{\rm vir}^{0.94\pm0.06}$ with very similar scatter to the correlation~(\ref{eq:jeans_vir}).

Comparing the virial and Schwarzschild $M/L$ estimates we can provide a direct `calibration' of the virial mass, and $M/L$ estimator (which are often used only in a relative sense):
\begin{equation}
(M/L)_{\rm vir}=\beta\; R_{\rm e}\sigma_{\rm e}^2/(L\, G).
\label{eq:virial}
\end{equation}
The best fitting scaling factor is $\beta=5.0\pm0.1$. We can compare this value with the predictions from simple dynamical models, as done by a number of previous authors \citep[e.g.][]{mic80}. For this we computed the theoretical predictions for $\beta$ from spherical isotropic models described by the S{\'e}rsic profile $R^{1/n}$, for different values of $n$. The computation was performed using high-accuracy MGE fits to the S{\'e}rsic profiles, obtained with the routines of \citet{cap02}. From the fitted MGE models, which reproduce the profiles to better than 0.05\%, the projected $\sigma$ values can be computed with a single one-dimensional numerical integration. The projected luminosity-weighted $\sigma$ was then integrated within a circular aperture of radius $R_{\rm e}$ to compute $\sigma_{\rm e}$ which is needed to determine the scaling factor $\beta$. In the range $n=2$--10 the predicted $\beta$ parameter is approximated to better than 3\% by the expression
\begin{equation}
\beta(n) = 8.87  - 0.831\; n + 0.0241\; n^2.
\label{eq:sersic}
\end{equation}
\citep[cf.][]{ber02}. The precise value predicted for a $R^{1/4}$ profile is $\beta=5.953$ (the value becomes $\beta=5.872$ with a BH of 0.14\% of the galaxy mass as in \citealt{har04}), while the observed value of $\beta\approx5.0$ would correspond to a S{\'e}rsic index $n\approx5.5$. However the predictions of equation~(\ref{eq:sersic}) only apply in an idealized situation and do not take into account the details in which $(M/L)_{\rm vir}$ is measured from real data and the fact that galaxies are not simple one-component isotropic spherical systems. From our extended photometry (\refsec{sec:phot}) we measured the $n$ values for the 25 galaxies of our sample by fitting the observed radial surface brightness profiles. The derived S{\'e}rsic indices span the whole range $n=2$--10 and will be presented in a future paper. From the observed variation in the profiles $\beta$ should be expected to vary by a factor $\sim2.5$ according to the idealized spherical model. In practice we find no significant correlation (linear correlation coefficient $r\approx-0.13$) between the measured $\beta$ (the value required to make $(M/L)_{\rm Schw}=\beta R_{\rm e}\sigma_{\rm e}^2/(L\, G)$) and the one predicted by equation~(\ref{eq:sersic}). This shows that the idealized model is not a useful representation of reality and cannot be used to try to improve the $(M/L)_{\rm vir}$ estimates. An investigation of the interesting question of why the $\beta$ parameter appears so constant in real galaxies and with realistic observing conditions goes beyond the scope of the present paper.

The results of this Section show that the simple virial estimate of $M/L$, and correspondingly of galaxy mass, is virtually unbiased, in the sense that it produces estimates that follow a nearly one-to-one correlation with the $M/L$ computed from much more `expensive' dynamical models. This result has implications for high redshift studies, as it shows e.g.\ that one can confidently use the virial estimator to determine masses at high redshift, where it is generally unfeasible to construct full dynamical models. It is important to emphasize however that this result strictly applies to virial measurements derived as we do in this paper, namely using `classic' determination of $R_{\rm e}$ and $L$ via $R^{1/4}$ growth curves, and with $\sigma_{\rm e}$ values measured in a large aperture.

\subsection{Comparison with previous results}

\begin{figure}
  \plotone{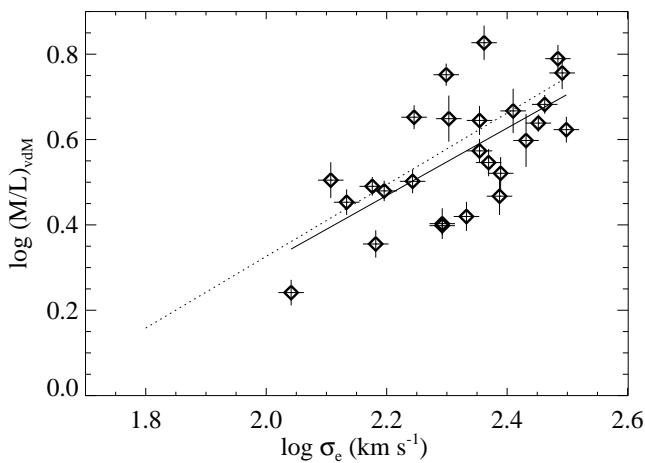}
    \caption{Same as in \reffig{fig:ml-sigma} for the $M/L$  measurements obtained by \citet{vdm91} on a different galaxy sample. His $R$-band $M/L$ values have been rescaled to the same SBF distances and zeropoint we use here and converted to the $I$-band assuming $(M/L)_R/(M/L)_I\approx1.27$. The solid line is the best-fitting relation to the data, while the dotted line shows for comparison relation~(\ref{eq:ml-sigma}).}
    \label{fig:ml-sigma_vdm91}
\end{figure}

To understand whether the tight $(M/L)$--$\sigma_{\rm e}$ correlation is really due to the improved quality of the kinematical data and photometric data, we determined the observed scatter in the Johnson $R$-band $M/L$ determinations by \citet{vdm91}, who used a similar Jeans modeling method, on a different sample of bright elliptical galaxies (five in common with our sample). From his sample of 37 galaxies we extracted the 26 for which SBF distances by \citet{ton01} exist and we performed a fit as done in \reffig{fig:ml-sigma}. We used in the correlations his mean velocity dispersion $\bar{\sigma}$ and his ``improved'' $(M/L)_R^{imp}$ (his column [14] from Table 1 and column [18] of Table 2 respectively), which we rescaled to the SBF distances, and converted to the $I$-band assuming an $R$-band absolute magnitude of the Sun of $M_{R,\sun}=4.42$ mag and a typical colour for ellipticals of $R-I\approx0.6$, which implies $(M/L)_R/(M/L)_I\approx1.27$. As in \refsec{sec:corr} we assumed a 5\% error in $\sigma$, and assigned to the $M/L$ the errors in the SBF distance, quadratically coadded to a constant error as to make $\chi^2/\nu=1$. The best-fitting relation is shown in \reffig{fig:ml-sigma_vdm91} and has the form:
\begin{equation}
(M/L) = (3.54\pm0.17) \left(\frac{\sigma_{\rm e}}{200\; \kms}
                                                    \right)^{0.79\pm0.17}
\end{equation}
which is in good agreement with our equation~(\ref{eq:ml-sigma}), but has much larger uncertainties and an increased observed scatter of 28\%. This increased scatter must be attributed mainly to the differences in the quality of the data used by \citet{vdm91} and in the present paper. For the five galaxies (NGC~3379, NGC~3608, NGC~4374, NGC~4486, NGC~5846) in common with this paper we measured relative differences $\epsilon=1-(M/L)_{\rm vdM}/(M/L)_{\rm Cap}$ of 26\%, 14\%, -11\%, -1\%, -27\% respectively.

\subsection{Comparison with stellar population $(M/L)_{\rm pop}$}
\label{sec:population}

We have established that the dynamical $M/L$ can be accurately determined from our data and models, and that a tight correlation exists with $\sigma_{\rm e}$, which also explains the appearance of the FP. The next obvious question is whether the observed $M/L$ variations are mainly due to a change in the stellar population or to differences in the dark matter fraction in galaxies. A number of authors have addressed this question usually using simple dynamical models combined with stellar population models. Sometimes these investigations have produced outcomes that are not entirely consistent between each other \citep[e.g.][]{kau03,pad04,dro04}. Here we use accurate dynamical $M/L$ and $(M/L)_{\rm pop}$ estimates from single stellar population (SSP) models, similarly to \citet{ger01}.

\begin{figure}
	\plotone{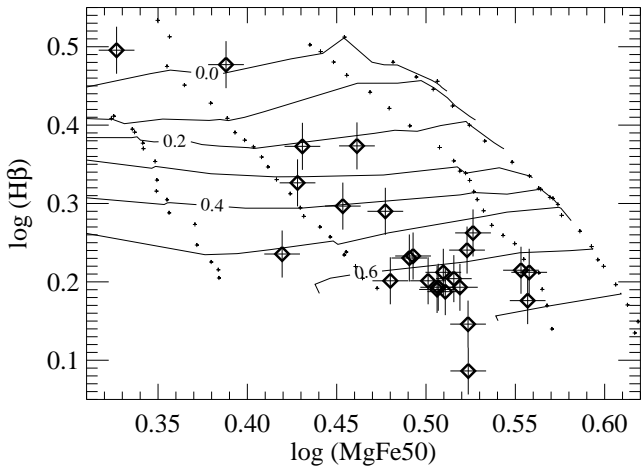}
	\caption{Observed values of the line strength index $\log [{\rm MgFe50}]$ versus $\log ({\rm H}\beta)$ for our galaxy sample. Overplotted are the contours showing the corresponding predictions of $\log(M/L)_{\rm pop}$ for the SSP models by VZ96 and VZ99, using the Salpeter IMF. The small crosses indicate the location where the model predictions are available, and from which the contours were linearly interpolated. Note the concentration of galaxies near the contour level $\log(M/L)_{\rm pop}=0.6$. The contours appear nearly horizontal in this plot, and the same is true using the \citet{kro01} IMF. The two galaxies with the ${\rm H}\beta$ value lower than any model prediction (M87 and NGC~5846) may be contaminated by residual gas emission. Their $(M/L)_{\rm pop}$ was estimated after increasing ${\rm H}\beta$ to the level of the model predictions.}
	\label{fig:hbeta-mgfe50}
\end{figure}

Using the $(M/L)_{\rm pop}$ predictions for the SSP models\footnote{Available from http://www.iac.es/galeria/vazdekis/. We used the model predictions for the $(M/L)_{\rm pop}$, that were updated in February 2005 to take into account the contribution of the remnants and the mass loss during the latest phases of the stellar evolution. The adopted lower and upper mass cutoffs for the IMF are 0.01 and 120 $M_\odot$ respectively, whereas the faintest star is assumed to be 0.09 $M_\odot$.} by \citet[hereafter VZ96]{vaz96}, and adopting as initial reference the \citet{sal55} stellar initial mass function (IMF), we derived $(M/L)_{\rm pop}$ from the \sauron\ Mg\,$b$, Fe5015 and H$\beta$ line-strength indices\footnote{The values were taken from Table~5 of Paper~VI after removing the applied Lick/IDS offsets listed in Table~3 of that paper.} \citep[Paper~VI]{kun06}.
To minimize the uncertainties in the absolute calibration of the line-strength indices to model predictions, we determined the index predictions for different SSP ages and metallicities by measuring them directly on the flux calibrated model spectral energy distributions produced by VZ99 (see discussion in \citealt{kun02}). We hereby circumvent the use of the so-called Lick fitting functions which can only be used in conjunction with uncertain offsets accounting for differences in the flux calibration between models and observations. The model spectra of VZ99 were broadened to the Lick resolution using the same procedure we adopted for the \sauron\ spectra. In order to minimize the influence of non-solar abundance ratios we derived the $(M/L)_{\rm pop}$ estimates from a [MgFe50] vs H$\beta$ diagram (\reffig{fig:hbeta-mgfe50}). The index $[{\rm MgFe50}]=(0.45\times {\rm Mg}b+{\rm Fe5015})/2$ is defined such that it is largely insensitive to abundance ratio variations, while H$\beta$ is only weakly dependent on it \citep{tra00,thod04}. For each galaxy, the line-strength measurements were extracted from the same high signal-to-noise \sauron\ spectrum from which the $\sigma_{\rm e}$ were derived, i.e. by luminosity-weighting all the spectra within $R_{\rm e}$ (\refsec{sec:kinematics}).

To test the robustness of the $(M/L)_{\rm pop}$ predictions we repeated the analysis of \reffig{fig:hbeta-mgfe50} using the models of \citet[BC03]{bru03}, and those by \citep[TMB03]{thod03} and \citet{mar05}, adopting the Salpeter IMF. The absolute values of the $(M/L)_{\rm pop}$ agree to within 20\% between the different models over the line-strength index range relevant for our galaxies. However the detailed agreement is better between the TMB03 and VZ96 models, while the predictions of BC03 tend to be noisier. In this paper we choose to use the VZ96 and VZ99 models because for them predictions of the spectral energy distribution at moderate resolution are also available which is helpful to minimize calibration issues.

\begin{figure}
	\plotone{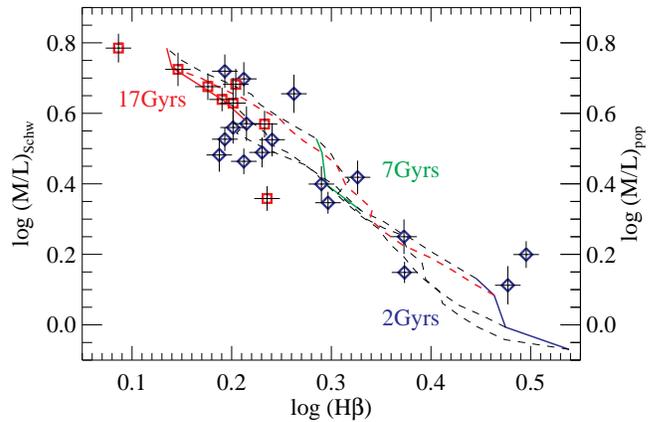}
	\caption{Observed values of the dynamical $M/L$ versus the observed line-strength index $\log ({\rm H}\beta)$. There is a clear inverse correlation between the two quantities, however the scatter is significant, and cannot be explained by measurement errors alone. The red squares and the blue diamonds indicate slow and fast rotating galaxies respectively, as defined in Table~\ref{tab1}. The coloured lines show the predictions of $\log (M/L)_{\rm pop}$ versus $\log ({\rm H}\beta)$ from the SSP models of VZ96 and VZ99, using a Salpeter IMF. The blue, green and red solid lines correspond to a model age of 2, 7 and 17 Gyr respectively. The dashed lines indicate a metallicity $[M/H]=[-0.68,-0.38,0.00,0.20]$, the red dashed line being solar metallicity $[M/H]=0$. The difference in the labelling of the left and right axes is meant to emphasize the fact that along the vertical axis we are comparing two different quantities: the dynamical or total $M/L$, and the $(M/L)_{\rm pop}$ of the stellar population alone. All the population model lines essentially overlap and define a tight relation between the two quantities, which is generally consistent with the measured trend. However the fact that some of the measured total $M/L$ lie below all of the population model predictions shows that the adopted IMF is unphysical. Adopting a Kroupa IMF the $(M/L)_{\rm pop}$ model predictions decrease by $\sim30\%$ ($\Delta\log(M/L)\sim 0.16$) and all the observed $M/L$ lie above or on top of the stellar population models. This indicates that a Kroupa IMF is consistent with the observations for all the galaxies in our sample.}
	\label{fig:ml-hbeta}
\end{figure}

We find that for single-burst stellar populations, the contour levels of constant $(M/L)_{\rm pop}$ in \reffig{fig:hbeta-mgfe50} tend to be nearly horizontal, in the sense that $(M/L)_{\rm pop}$ is essentially a function of H$\beta$ alone. This implies that, under the model assumptions, if the variations in the dynamical $M/L$ are driven by the variation in the stellar population, a good correlation should exist between $M/L$ and H$\beta$. This is tested in \reffig{fig:ml-hbeta}, where we plot the measured $M/L$ and H$\beta$ measurements and compare them to the $(M/L)_{\rm pop}$ versus H$\beta$ predictions, for a set of models with a large spread in age (2--17 Gyr) and metallicity ($[M/H]=[-0.68,-0.38,0.00,0.20]$). As expected from \reffig{fig:hbeta-mgfe50} the envelope of the different model curves traces a tight relation, nearly linear in logarithmic coordinates (this is less true if one also considers models with more extreme $[M/H]$ values). The model relation follows the same trend as the data, suggesting that the variation of the stellar population is indeed an important factor in determining the observed variations of the $M/L$. The span of $M/L$ from 1--6 mainly reflects differences in the luminosity-weighted age of the stellar population. The observed age trend is consistent with evidences of downsizing \citep{cow96} from high redshift studies \citep[e.g.][]{dis05,tre05,vdw05}.

The accuracy of our $M/L$ determinations allows us to go beyond the general agreement, to detect significant deviations, and to exclude a simple one-to-one relation between the total $M/L$ given by the dynamical models and the $(M/L)_{\rm pop}$ predicted by the stellar population models. In particular, for some galaxies, the measured total $M/L$ is lower than any of the model predictions with the Salpeter IMF. For this reason, with the above caveats about the absolute scaling of the $(M/L)_{\rm pop}$ values, and if we assume the IMF to be the same for all galaxies, we have to reject the Salpeter IMF (and any IMF with larger slope) as it gives unphysical results. \citet{kro01} has measured the IMF and constrained the shape below one solar mass with the result that there are fewer low mass stars than indicated by the Salpeter law. The effect of using the Kroupa IMF is just to decrease all the $(M/L)_{\rm pop}$ model predictions by $\sim30\%$ ($\Delta\log(M/L)\sim 0.16$). With this IMF, none of the dynamical $M/L$ values lie below the relation defined by the model predictions, within the measurement errors. Although we cannot independently confirm the correctness of this IMF we adopt it for the purposes of the following discussion.

\begin{figure}
	\centering\includegraphics[width=0.75\columnwidth]{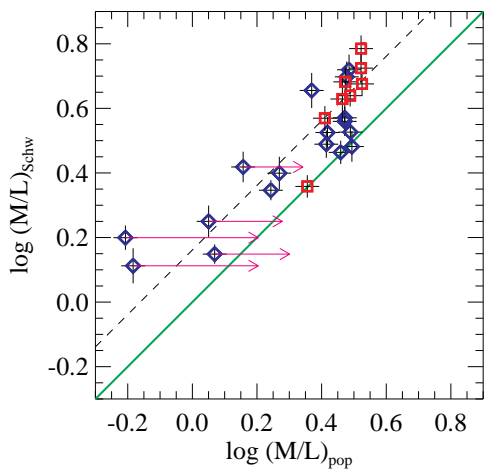}
	\caption{Dynamical (i.e.\ total) $M/L$ from the Schwarzschild modeling as a function of $(M/L)_{\rm pop}$ using the SSP models of VZ96 and VZ99, with a \citet{kro01} IMF. The red squares and the blue diamonds indicate the slow and fast rotating galaxies respectively, as defined in Table~\ref{tab1}. The thick green line indicates the one-to-one relation. All galaxies have $(M/L)_{\rm pop}\la (M/L)$ within the errors, but the total and stellar $M/L$ clearly do not follow a one-to-one relation. Dark matter is needed to explain the differences in $M/L$ (if the IMF is not varying).  The magenta arrows show the variation in the estimated $(M/L)_{\rm pop}$ for the youngest galaxies (luminosity-weighted age $\la7$ Gyr), if a two-population model is assumed (see text for details). The $(M/L)_{\rm pop}$ of the young galaxies would move closer to the one-to-one relation. Adopting the Salpeter IMF all the values of $(M/L)_{\rm pop}$ would increase by $\Delta\log(M/L)\sim 0.16$. This can be visualized by shifting the one-to-one relation to the position of the dashed line. In this case a number of galaxies would have $(M/L)_{\rm pop}>(M/L)$ and this implies that the Salpeter IMF is unphysical (consistent with \reffig{fig:ml-hbeta}).}
	\label{fig:ml-mlpop}
\end{figure}

A more direct, but more model dependent, way of looking at the relation between the stellar population and the $M/L$ variations, is to compare $M/L$ with $(M/L)_{\rm pop}$ predicted using the Kroupa IMF (\reffig{fig:ml-mlpop}). The similarity of this figure with \reffig{fig:ml-hbeta}, which was obtained from the data alone, shows that the structure seen in the plot is robust, and does not come from subtle details in the SSP predictions. Again the main result is a general correlation between the total $M/L$ and the stellar population $(M/L)_{\rm pop}$, consistent with \citet[see their Fig.~14]{ger01}, albeit with a smaller scatter. The relatively small scatter in the correlation indicates that the IMF of the stellar population varies little among different galaxies, consistent with the results obtained for spiral galaxies by \citet{bel01}. All galaxies have $(M/L)_{\rm pop}\la (M/L)$ within the errors, but the total and stellar $M/L$ clearly do not follow a one-to-one relation (green thick line). Unless this  difference between $M/L$ and $(M/L)_{\rm pop}$ is related to IMF variations between galaxies, it can only be caused by a change of dark matter fraction within\footnote{This statement is not entirely rigorous, in fact, as shown in \refsec{sec:dark-matter}, the presence of a dark matter halo can increase the $M/L$ measured within $R_{\rm e}$ in a way that is not simply related to the amount of dark matter within that radius.} one $R_{\rm e}$. The inferred median dark matter fraction is 29\% of the total mass, broadly consistent previous findings from dynamics \citep[e.g.][]{ger01,thoj05} and gravitational lensing \citep[e.g.][]{tre04,rus05}, implying that early-type galaxies tend to have `minimal halos' as spiral galaxies \citep{bel01}. The large number of galaxies in our sample with a similar value of $\log(M/L)_{\rm pop}\sim0.5$, and the clear evidence for a variation in the dark matter fraction in those galaxies is consistent with the result of \citet{pad04}. However our results are not consistent with a constancy of the $(M/L)_{\rm pop}$ for all galaxies, and a simple dark-matter variation, as we clearly detect a small number of young galaxies with low $(M/L)_{\rm pop}$ and a correspondingly low dynamical $M/L$.
Adopting the Salpeter IMF all the values of $(M/L)_{\rm pop}$ would increase by $\Delta\log(M/L)\sim 0.16$. This can be visualized by moving the one-to-one relation to the position of the dashed line. In this case a number of galaxies would have $(M/L)_{\rm pop}>(M/L)$, implying that the Salpeter IMF is unphysical (as we inferred from \reffig{fig:ml-hbeta}).

This analysis has been carried out in the context of SSP models. We now consider a few caveats arising from this methodology. Low-level secondary star formation can alter the values of the $(M/L)_{\rm pop}$ derived: essentially we underestimate $(M/L)_{\rm pop}$ for these galaxies. This effect is expected to influence mainly the youngest galaxies, with the lowest $(M/L)_{\rm pop}$. To estimate the importance of this effect we experimented with combinations of two SSP models, considering a young component on top of an older solar-metallicity one. In \reffig{fig:ml-mlpop} we indicate with an arrow the variation in the estimated $(M/L)_{\rm pop}$, for the five youngest galaxies (luminosity weighted age less than 7 Gyr), {\em assuming} the stellar population is composed by 90\% (in mass) by an old 12 Gyr population with solar metallicity and by 10\% by a younger population, whose age and metallicity is allowed to vary to reproduce the observed line strengths. We find that the artificial underestimation of the lowest $(M/L)_{\rm pop}$ values can easily explain why these galaxies appear to have a high fraction of dark matter, contrary to the general trend. The adopted models use element abundances in the solar ratios, yet we know that many of our sample galaxies have enhanced ratios of Mg/Fe. Unfortunately, there are no models available which predict $(M/L)_{\rm pop}$ as a function of abundance ratios. A possible influence on our results remains to be explored.

Given our assumptions about the IMF, we conclude the following: (i) There is a reasonable correlation between the dynamical and stellar population estimates of $M/L$ in our sample. (ii) At low stellar $M/L$ the overall relation is driven by age. (iii) For old galaxies there is a trend in dark matter fraction within one $R_{\rm e}$, from zero ($(M/L)\simeq 3$) to about 30\% dark matter contribution for galaxies with the highest $M/L$ in our sample ($(M/L)\simeq 6$).

\section{Discussion}
\label{discussion}

\subsection{A second parameter in the $M/L$ variations?}
\label{sec:rotators}

In this Section we investigate whether the variations in the $M/L$ of the galaxies, apart from a main dependence on the galaxy $\sigma_{\rm e}$ (and closely related quantities like mass and luminosity), are also dependent on other global observables, in particular on the galaxy kinematics, morphology and environment.

As mentioned in \refsec{sec:intro}, a number of studies over the past twenty years have suggested that early-type galaxies display a dichotomy between generally massive, slowly-rotating, metal-rich galaxies and rotationally-supported, less massive, metal-poor galaxies, which suggests a difference in their formation scenarios \citep[e.g.][]{dav83,fab97}. These differences in galaxy structure led \citet{kor96} to propose a physically-motivated revision to the classic and still broadly used (e.g.\ RC3) visual classification scheme by \citet{hub36}, based on photometry of E and S0 galaxies, to take the findings from the kinematics into account. The fainter and faster rotating galaxies also tend to show steeper luminosity profiles than the brighter ones \citep{fab97} and there is indication that they may all contain disks \citep{lau05}. The \sauron\ kinematical maps presented in Paper~III also show two general velocity field morphologies: galaxies with a clear sense of rotation, with only small kinematic twists; and galaxies with little or no rotation, or with strong kinematic misalignments. The relation between all these different galaxy properties is still not fully understood, and it will be addressed in a future paper, for the full \sauron\ representative sample.

Here we limit ourselves to studying the effect on the $M/L$ of a single, hopefully representative, parameter: the galaxy kinematics. A classic and simple way to quantify the differences in galaxy kinematics is by using the anisotropy diagram \citep{bin78}, which relates the observed flattening $\varepsilon$ of a galaxy to the ratio between the ordered and random rotation of the stars ($V/\sigma$). The formalism was recently updated by \citet{bin05b} for use with integral-field kinematics.  The $(V/\sigma)-\varepsilon$ diagram has often been used to provide a separation between fast rotating and slow rotating early-type galaxies \citep[e.g.][]{geh03}. A limitation of the diagram to characterize the kinematics of a galaxy is that it contains no spatial information, so that it cannot distinguish between a fast rotating nuclear component and a global galaxy rotation. To overcome this limitation we introduced a new {\em quantitative} classification scheme for early-type galaxies which is based on an approximate measure of the specific angular momentum of galaxies from integral-field stellar kinematics. In this work we adopt that classification to define the galaxies in our sample with significant angular momentum per unit mass, which we call the ``fast-rotators'' and the ones with negligible amount of specific angular momentum, which we define ``slow-rotators'' (Table~\ref{tab1}). The distinction between the two classes of galaxies can be easily seen on the kinematic maps of Paper~III. More details on the classification will be given in Emsellem et al.\ (in preparation).

\begin{figure}
	\centering\includegraphics[width=0.8\columnwidth]{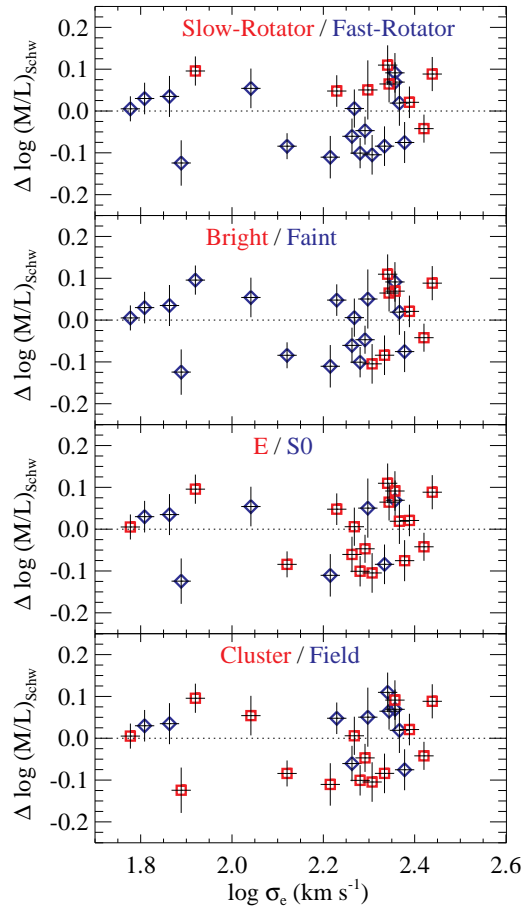}
	\caption{Residuals in the $(M/L)$--$\sigma$ correlation of \reffig{fig:ml-sigma}. The values shown are the same in the four panels, but the meaning of the red and blue symbols is different in each panel. {\em First Panel:} the different galaxies are coloured according to their kinematics, using the classification in Table~\ref{tab1}. Namely the blue diamonds are fast-rotator galaxies, while the red squares are the slow-rotator galaxies.  {\em Second Panel:} the blue diamonds and red squares are the galaxies respectively fainter and brighter than $M_K=-24$. {\em Third Panel:} here blue diamonds are S0 galaxies and red squares are E galaxies, as defined in RC3. {\em Fourth Panel:} here blue diamonds are field galaxies, while red squares are cluster galaxies (as defined in Paper~II). The K-S probability that the observed distribution of the residuals, for each pair of classes of objects, is drawn from the same distribution is 2\%, 32\%, 74\% and 10\% from the top to the bottom panel respectively. The only significant difference in $M/L$ is thus observed between the fast-rotators and the slow-rotators, while we do not detect any significant difference as a function of luminosity or morphological classification. There is some very marginal evidence for a difference in $M/L$ as a function of environment.}
	\label{fig:residuals}
\end{figure}

In the first (top) panel of \reffig{fig:residuals} we show the distribution of the residuals in the $(M/L)$--$\sigma$ correlation for the fast-rotators and slow-rotators, as defined above. There appears to be a tendency for the slow-rotators to have a higher $M/L$ than the fast-rotators, at given $\sigma_{\rm e}$. This difference in the distribution of the residuals between the two classes of objects can be quantified using the Kolmogorov-Smirnov (K-S) test \citep[e.g.][Section~14.3]{pre92}. We find that the two distributions are significantly different, the probability that they are drawn from the same distribution being 2\%. The same probability is found using a completely different estimator, the median statistics \citep{got01}, counting the number of galaxies above and below the best-fitting relation.

A similar difference between the $M/L$ of the slow-rotators and fast-rotators is seen with respect to the $(M/L)_{\rm pop}$ in \reffig{fig:ml-mlpop} (or \reffig{fig:ml-hbeta}), in the sense that the slow rotating galaxies tend to have the largest $M/L$ at any given $(M/L)_{\rm pop}$ (or H$\beta$). If one excludes from the comparison the five youngest galaxies, and only considers the older galaxies for which the $(M/L)_{\rm pop}$ is likely to be reliable, one finds that 7/8 of the slow rotating galaxies lie on top or above the dashed line (which roughly corresponds to a $\sim30\%$ contribution of dark matter within $R_{\rm e}$), while only 3/12 of the fast rotating galaxies lie above the same line. In this case the probability that this difference in $M/L$ is due to a random fluctuation is less than 1\%. The only slow rotator without evidence for dark matter is NGC~4458, which appears peculiar also for being the only low-luminosity and low-$\sigma_{\rm e}$ slow rotator.

In the second panel of \reffig{fig:residuals} we show the dependence of the residuals of the $(M/L)$--$\sigma$ relation on the galaxy luminosity. We separated the sample into the galaxies brighter and fainter than the K-band luminosity $M_K=-24$. This ad hoc luminosity value was chosen to have the same number of objects in the two classes as in the comparison of the top panel between fast and slow rotating galaxies. We do not detect any significant difference in the distribution of the residuals of the bright and faint galaxies. The K-S probability that the residuals come from the same distribution is in this case 32\%.

This finding that the slow rotating galaxies, which are more common among the most massive ones (\reffig{fig:residuals}), have a higher $M/L$ than the faster rotating and generally less massive ones, is reminiscent of recent findings, derived using completely different techniques, either from the dynamics of the planetary nebulae \citep{rom03,nap05} or from the analysis of gravitationally-lensed quasars \citep{fer05}, indicating smaller fractions of dark matter among faint ellipticals, compared to more massive ones. However, if our result is correct, it would indicate that the dark matter fraction is more closely linked to the galaxy dynamics than to the mass alone.

In the third panel of \reffig{fig:residuals} we compare the distribution of the residuals in the $(M/L)$--$\sigma$ relation for the morphologically classified E and S0 (from RC3). Consistent with previous studies \citep[e.g.][]{jor96} we find no difference in the $M/L$ distribution for the two samples. The K-S probability that the two $M/L$ distributions are the same is in fact 74\%. The lack of any dependence on galaxy morphology may just reflect the limitations of the classic E/S0 visual classification, with respect to more physically-motivated ones, which also consider the galaxy kinematics \citep[e.g.][]{kor96}.

Finally in the fourth (bottom) panel of \reffig{fig:residuals} we test whether the $M/L$ shows a different distribution for the subsample of galaxies in clusters and in the field (defined as in Paper~II). Hierarchical galaxy formation scenarios \citep{dia01} predict that the galaxies in the field should have a significantly {\em lower} $M/L$ than the galaxies in clusters, nearly independently of redshift. From our sample we find some very marginal difference between the two distributions of residuals, the K-S probability that the two distributions are the same being 10\%. This small difference goes in the opposite direction than the model predictions, in the sense that the field galaxies have a marginally higher $M/L$ than the cluster galaxies. This is consistent with recent findings from high-redshift studies, which show a similar evolution for the $M/L$ of cluster and field galaxies \citep[e.g.][]{vdv03,vdw05}. Together with the existence of the correlations we discuss in this paper, this result shows that the physical properties of the individual galaxies play a more important role in their formation and evolution than the environment.

\subsection{Origin of the $(M/L)$--$\sigma$ correlation}

The $(M/L)$--$\sigma_{\rm e}$ correlation, which we studied in detail in this paper, is part of the remarkable series of tight relations of the $\sigma_{\rm e}$ in early-type galaxies and spiral bulges with (i) luminosity, (ii) colours, (iii) metallicity and more recently also (iv) BH mass and (v) mass of the dark halo \citep{fer02}. Together with these regularities, early-type galaxies are characterized by dichotomies, between the most massive giant ellipticals on one side, which are red, metal-rich, slowly-rotating, and have shallow photometric profiles, and the fainter objects on the other side, which tend to be bluer, metal-poor, are dominated by rotation, have cuspy profiles and may contain disks \citep{dav83,kor96,fab97,lau05}.  One way to explain the observed trends is with the standard hierarchical scenario, where galaxies are assembled from smaller building blocks. In this picture the importance of gas during the merger process declines with increasing galaxy mass, due to the transformation of gas into stars, while galaxies are assembled. The role of the gas dissipation in the last merger could explain the observed differences in the characteristics of early-type galaxies \citep{ben92}. However it was realized some time ago that an apparently more successful way of explaining both the regular relations and the dichotomies is to assume some form of feedback process \citep{dek86}, which must be able to stop the star formation and the chemical evolution by heating and expelling the gas from the galaxy potential well (which is related to the observed $\sigma$).

Until recently the most natural candidate feedback generator was the explosion of supernovae. But the last few years of BH studies have started to revolutionize this picture. Just five years ago the $M_{\rm BH}$--$\sigma$ relation first suggested a close connection between the formation of the BH and of the galaxy stellar component. Now a number of lines of evidence seem to converge towards a scenario in which the BH is the main actor in the feedback process which shapes galaxy evolution \citep[e.g.][]{sil98,gra04,bin05a,ost05,loe05,spr05}. In this picture the gas accumulates in the dark matter potential well, feeding the BH, forming stars and enriching the medium with metals, until the BH grows to the point where its powerful jets heat and expel the gas from the galaxy. This scenario naturally explains a close connection of $\sigma$ with the BH mass \citep{dim05}, as well as with the asymptotic circular velocity (related to the dark halo mass) and the $M/L$ (related to both the halo mass and the metal enrichment).

The most obscure part of this picture is the link between the dark and the visible matter, both in the centres and the outskirts of early-type galaxies. It is remarkable that, thirty years after the first evidence of dark matter in spiral galaxies, a clear evidence of dark matter in early-type galaxies is still missing. Although a number of isolated studies from stellar dynamical models, X-ray and HI gas kinematics, or gravitational lensing indicate, not surprisingly, that early-type galaxies are also embedded in massive dark halos, other results have challenged this picture, suggesting that some galaxies may not contain dark matter after all. In this context our work helps to constrain the elusive dark matter contribution in the central regions of galaxies. The accurate $M/L$ determinations in galaxy centres, compared with $M/L$ derived by kinematical tracers at large radii, are also needed to measure dark matter profiles.

Progress on the investigation of the $(M/L)$--$\sigma$ relation can be made using: (i) larger galaxy samples, to reliably determine the intrinsic scatter and the significance of the apparent connection between the dynamics and the dark matter content in galaxies; (ii) more accurate distances (e.g.\ using the SBF method from the ACS Virgo Cluster Survey [\citealt{cot04}], which includes some galaxies in our sample); (iii) integral-field observations of samples of galaxies in clusters at intermediate distance, where the kinematics can still be resolved, but galaxies can be assumed to lie at the same distance.

\subsection{Caveats}

The galaxy sample we use in this paper was extracted from the \sauron\ representative sample of E/S0 galaxies in Paper~II, and in addition includes M32 to explore the small mass regime. As described in \refsec{selection}, the main selection criteria were the availability of HST photometry and SBF distances for the galaxies, but we also excluded some obvious bars (a similar selection affects the well known $M_{\rm BH}$--$\sigma$ relation). Although we have no indication that these selection criteria should bias our results in any specific way, they make the sample not necessarily representative any more. The fact that we can recover the FP correlations suggests that our small sample does represent the E/S0 population, but it is clear that increasing the galaxy modeling sample in a statistically selected way would be beneficial for future studies.

The slow rotating galaxies that we discussed in \refsec{sec:rotators} are generally thought to be triaxial \citep[e.g.][]{dez91}. If this is the case our axisymmetric models will only provide a first order approximation to the true orbital distribution of the galaxies, and the $M/L$ may not be correctly recovered. A significant triaxiality however may be expected to increase the observed scatter in the $(M/L)$--$\sigma$ relation, which is instead quite small, arguing against strong triaxiality of the slow rotating galaxies in our sample. This idea is also supported by the fact that these galaxies all appear quite round in projection. In any case it is not obvious that axisymmetric models of a sample of possibly triaxial galaxies, seen at random orientations, should produce a {\em positive} bias in the inferred $M/L$. A detailed test using triaxial galaxy models \citep{ver03} goes beyond the scope of the present paper.

In our models we assumed a constant $M/L$ throughout the galaxies and in \refsec{sec:dark-matter} we discussed the possible implications of this assumption. It is possible for the dark matter profile to vary systematically as a function of galaxy mass. As a result our constant $M/L$ approximation would not be equally accurate for the galaxies in the different mass ranges. This effect could also produce a bias on the measured $M/L$, which is however difficult to quantify, given the limited information on the dark matter profiles.

The result of \refsec{sec:dark-matter} shows, however, that it is not easy to `hide' dark-matter in the centre of galaxies. For this reason the tight $(M/L)$--$\sigma$ correlation implies that (i) either dark matter is unimportant in galaxy centres or (ii) dark matter is closely related to the luminous matter. In fact a dark matter halo that dominates the central regions more than the one considered (e.g.\ the steeper profiles found in the numerical simulations of \citealt*{nav96}), would easily erase the correlation derived from the $M/L$ of our models, unless the halo properties are closely linked to the stellar component. A similar conclusion was reached in a study of the tightness of the FP by \citet{bor03}.

\section{Conclusions}
\label{conclusion}

We measured the $M/L$ in the central regions, for a sample of 25 early-type galaxies with \sauron\ integral-field kinematics, using both two-integral Jeans models and three-integral Schwarzschild models. The galaxy sample was extracted from the \sauron\ survey (Paper~II) and spans over a factor of a hundred in mass. To this sample we added the galaxy M32 to explore the low-mass range. The models are constrained by \sauron\ kinematics, to about one effective (half-light) radius $R_{\rm e}$, and HST/WFPC2 $+$ MDM ground-based photometry in the $I$-band.  We studied the correlations of the $M/L$ with global observables of the galaxies and found a sequence of increasing scatter when the $M/L$ is correlated to: (i) the luminosity-weighted second moment of the velocity inside the half-light radius $\sigma_{\rm e}$, (ii) the galaxy mass, (iii) the $K$-band luminosity and (iv) the $I$-band luminosity. These correlations suggest that the $M/L$ depends primarily on galaxy mass, but at a given mass it is a function of $\sigma$, which in turn is related the galaxy compactness.  For our galaxy sample the $(M/L)$--$\sigma_{\rm e}$ relation has an observed scatter of 18\%  and an inferred intrinsic scatter of $\sim13\%$.

We compared the $M/L$ derived from the dynamical models, which does not depend on any assumption of spatial or dynamical homology, with the classic predictions for the $(M/L)_{\rm FP}$ derived from the Fundamental Plane, which fully depends on the assumptions of homology and virial equilibrium.  The slope of our observed correlations of $M/L$ with luminosity, mass and $\sigma$, can account for $\sim90\%$ of the tilt of the FP.  We also compared directly the $M/L$ from the models with the virial predictions for our own sample, and found a relation $(M/L)\propto(M/L)_{\rm vir}^{1.08\pm0.07}$, which confirms, with smaller uncertainties, the results from the FP comparisons. This also shows that, when the virial parameters $R_{\rm e}$ and $L$ are determined in the `classic' way, using growth curves and the $\sigma_{\rm e}$ is measured in a large aperture, the virial mass is a reliable and unbiased estimator of galaxy mass. This has implications for high redshift studies, where the construction of general dynamical models is unfeasible. The best-fitting virial relation has the form $(M/L)_{\rm vir}=(5.0\pm0.1)\times R_{\rm e}\sigma_{\rm e}^2/(L\,G)$, in broad agreement with simple theoretical predictions.

These correlations raise the question whether the observed $M/L$ variations are primarily due to differences in the $(M/L)_{\rm pop}$ of the stellar population or differences in the dark-matter fraction. To test this we compared our dynamical $M/L$ with simple estimates of the $(M/L)_{\rm pop}$, derived from the observed line-strengths indices of our galaxies. We find that the $M/L$ generally correlates with the $(M/L)_{\rm pop}$, indicating that the variation in the stellar population, mainly due to a variation in the luminosity-weighted age, is an important factor in driving the observed $M/L$ variation. The relatively small scatter in the correlation indicates that the IMF of the stellar population varies little among different galaxies, consistent with results obtained for spiral galaxies by \citet{bel01}. The small scatter also argues against the cuspy dark matter density profiles predicted by numerical simulations \citep{nav96}.

The accuracy of our $M/L$ determinations allows us to detect significant deviations from a one-to-one correlation, which must be due to either variations of the IMF among galaxies, or to variations in the dark matter fraction within $R_{\rm e}$. In the latter case the median dark matter fraction for our sample is found to be $\sim30\%$ of the total mass, broadly consistent with earlier findings from dynamics \citep[e.g.][]{ger01,thoj05} and gravitational lensing \citep[e.g.][]{tre04,rus05}. We find some evidence for the variation in $M/L$ to be related to the dynamics of the galaxies. In particular the slow rotating galaxies in our sample, which are more common among the most luminous objects, tend to have a higher $M/L$ at given $(M/L)_{\rm pop}$ than the fast rotating and generally fainter galaxies. Assuming a constant IMF among galaxies, these results would suggest that the slow rotating massive galaxies have a higher ($\sim30\%$) dark matter fraction than the fast rotating galaxies. We speculate that this difference in $M/L$ indicates a connection between the galaxy assembly history and the dark halo structure.

\section*{Acknowledgments}

We thank Marijn Franx for very useful discussions, Anne-Marie Weijmans for careful reading of the manuscript and for commenting on the draft, and Franco Maschietto for support with the tests of the modeling code. We are grateful to Roeland van der Marel for important suggestions on our submitted paper. We thank Barbara Lanzoni, Luca Ciotti and Alexandre Vazdekis for useful comments.
The \sauron\ project is made possible through grants 614.13.003, 781.74.203, 614.000.301 and 614.031.015 from NWO and financial contributions from the Institut National des Sciences de l'Univers, the Universit\'e Claude Bernard Lyon~I, the Universities of Durham, Leiden, and Oxford, the British Council, PPARC grant `Extragalactic Astronomy \& Cosmology at Durham 1998--2002', and the Netherlands Research School for Astronomy NOVA.  RLD is grateful for the award of a PPARC Senior Fellowship (PPA/Y/S/1999/00854) and postdoctoral support through PPARC grant PPA/G/S/2000/00729. The PPARC Visitors grant (PPA/V/S/2002/00553) to Oxford also supported this work. MB acknowledges support from NASA through Hubble Fellowship grant HST-HF-01136.01 awarded by Space Telescope Science Institute, which is operated by the Association of Universities for Research in Astronomy, Inc., for NASA, under contract NAS~5-26555 during part of this work. MC acknowledges support from a VENI award 639.041.203 awarded by the Netherlands Organization for Scientific Research (NWO). JFB acknowledges support from the Euro3D Research Training Network, funded by the EC under contract HPRN-CT-2002-00305. This project made use of the HyperLeda and NED databases. Part of this work is based on data obtained from the ESO/ST-ECF Science Archive Facility. Photometric data were obtained (in part) using the 1.3m McGraw-Hill Telescope of the MDM Observatory.

{}

\appendix

\section{Constraints on inclination from two-integral models}
\label{sec:inc}

\begin{figure*}
  \includegraphics[width=\textwidth]{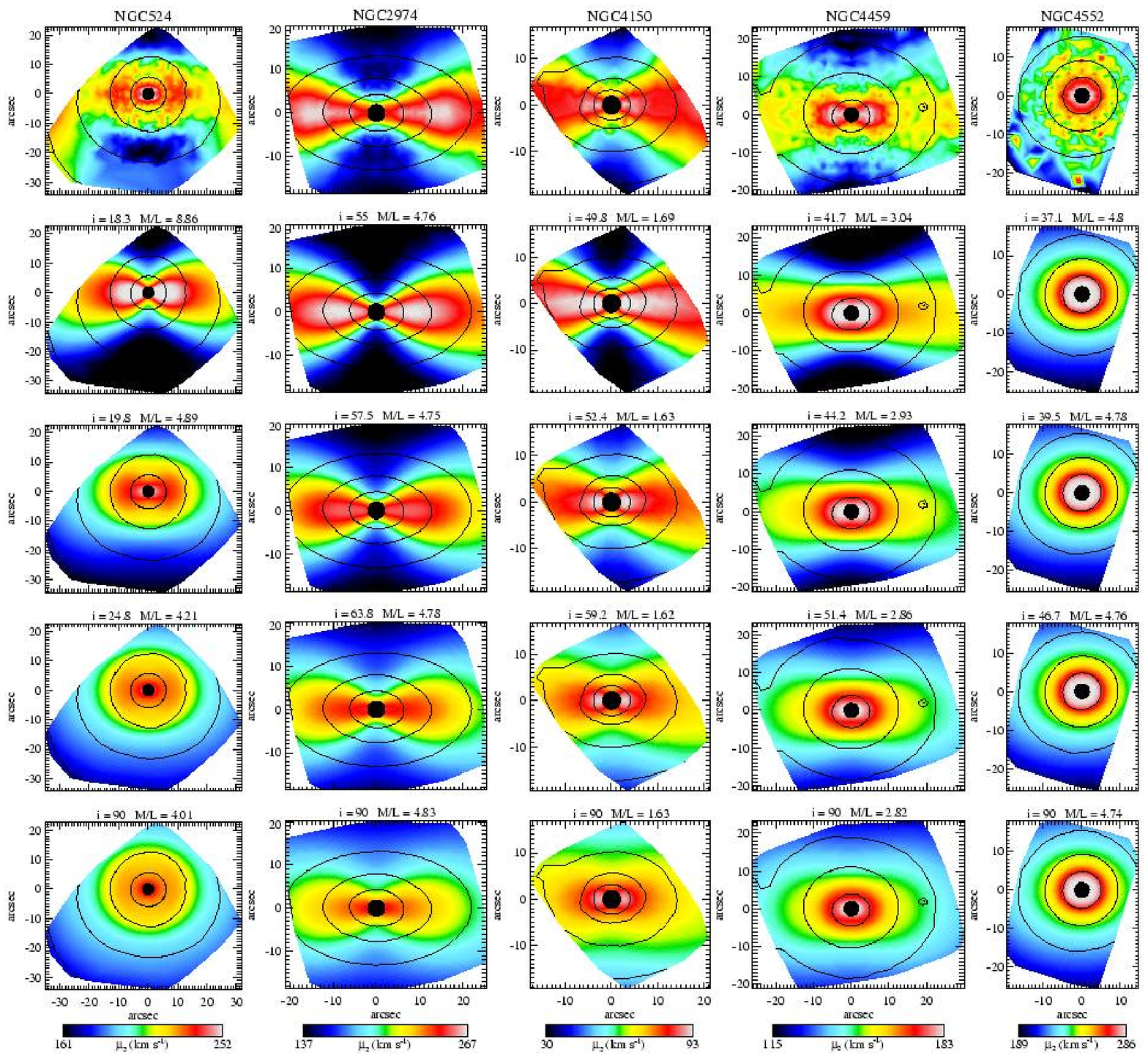}
    \caption{{\em Top Panels:} Bi-symmetrized and linearly interpolated map of the observed second velocity moment $\mu'_2\equiv\sqrt{V^2+\sigma^2}$, extracted from the \sauron\ stellar kinematics of the galaxies NGC~524, NGC~2974, NGC~4150, NGC~4459 and NGC~4552. The actual Voronoi bins in which the kinematics was measured are shown in Paper~III. The black central disk shows the region that was excluded from the following fits. {\em Subsequent Panels:} two-integral Jeans modeling of the unique projected second moments associated to the MGE mass density distribution, for different inclinations. Each plot is scaled to the best-fitting $M/L$ using equation~(\ref{eq:fit}). For every galaxy the lowest inclination corresponds to an intrinsic axial ratio $q=0.1$ of the flattest Gaussian component in the MGE model. Subsequent panels are equally spaced in $q$. The inclination of each model, and the corresponding best-fitting $M/L$ is written above each plot. The inclination independently inferred from the geometry of the dust or from the gas kinematics is $i\approx20^\circ,60^\circ,54^\circ,46^\circ$, for NGC~524, NGC~2974, NGC~4150 and NGC~4459 respectively. No strong constraint exists for the inclination of NGC~4552, but this giant elliptical is not expected to be flat (see \refsec{sec:jeans}), so it should be best fitted for $i\approx90^\circ$. For all the five galaxies the best fitting inclination derived from the Jeans models agrees with the expected value, as can also be seen by eye on these maps.}
    \label{fig:ngc524_maps}
\end{figure*}

In this Appendix we present the Jeans two-integral models of five galaxies in our sample, for which the inclination can be independently determined by non-dynamical arguments. The galaxies NGC~524 and NGC~4459 possess very regular dust disks, while NGC~4150 has a more disturbed disk, which are all clearly visible on the unsharp-masked HST/WFPC2 images \citep[Fig.~4 of][Paper~V]{sar06}. We fitted an ellipse to the dust morphology, and, assuming the disks are intrinsically circular, and in equilibrium in the equatorial plane of an oblate spheroid, we inferred an inclination of $i=20^\circ\pm5$, $i=54^\circ\pm5$ and $i=46^\circ\pm2$, respectively for NGC~524, NGC~4150 and NGC~4459. The inclination of the gas disk in NGC~2974 was carefully studied by \citet{kra05} who derived an inclination of $i=60^\circ\pm3$. Finally in the case of NGC~4552 there is no evidence of extended patchy dust absorption, but this giant elliptical is unlikely to be intrinsically very flat (see discussion in \refsec{sec:jeans}). We expect a broad range of possible inclinations, but low values should be excluded.

In \reffig{fig:ngc524_maps} we compare the maps of the observed projected second moment of the velocity $\mu'_2\equiv\sqrt{V^2+\sigma^2}$, as derived from the \sauron\ stellar kinematics, with the maps of the unique second velocity moment associated to the MGE density distribution, computed by solving the Jeans equations, at every inclination. To help the visual comparison between data and models, the observed $\mu'_2$ was bi-symmetrized by replacing the value in each bin with the average of the corresponding values in the four quadrants (computed using linear interpolation). Each plot is scaled to the best-fitting $M/L$ using equation~(\ref{eq:fit}). This same kind of fit was performed for all the 25 galaxies in the sample, to determine the inclination at which the $M/L$ of the Jeans and Schwarzschild models was measured. The Jeans models always provide a clear best fitting value for the inclination, which can also be easily recognized by eye. Starting from an edge-on model ($i=90^\circ$), the effect of lowering the inclination (while keeping the $M/L$ at the best fitting value) can be mainly described as a decrease of the second moments along the minor axis, accompanied by a corresponding increase along the major axis. For the five galaxies considered in this Appendix the best-fitting inclination derived from the Jeans models agrees with the expected values.

\section{MGE parameters of the sample}

\begin{table*}
\caption{MGE parameters for the deconvolved $I$-band surface brightness.}
\centering
\begin{tabular}{ccccccccccccc}
\hline
$j$ & $\log I'_j$ & $\log\sigma_j$ & $q'_j$  &  $\log I'_j$ & $\log\sigma_j$ & $q'_j$   & $\log I'_j$ & $\log\sigma_j$ & $q'_j$  & $\log I'_j$ & $\log\sigma_j$ & $q'_j$  \\
    & (L$_{\odot,I}$ pc$^{-2}$) & (arcsec) & & (L$_{\odot,I}$ pc$^{-2}$) & (arcsec) &   & (L$_{\odot,I}$ pc$^{-2}$) & (arcsec) & & (L$_{\odot,I}$ pc$^{-2}$) & (arcsec) &   \\
\hline
 & \multicolumn{3}{c}{NGC 221} & \multicolumn{3}{c}{NGC 524} & \multicolumn{3}{c}{NGC 821} & \multicolumn{3}{c}{NGC 3156} \\
\hline
    1 &   6.187 &  -1.762 &   0.790 &   4.336 &  -1.762 &   0.950 &   5.322 &  -1.526 &   0.706 &   6.103 &  -1.762 &   0.700 \\
    2 &   5.774 &  -1.143 &   0.741 &   3.807 &  -1.199 &   0.950 &   4.922 &  -0.954 &   0.703 &   5.236 &  -1.464 &   0.700 \\
    3 &   5.766 &  -0.839 &   0.786 &   4.431 &  -0.525 &   0.950 &   4.608 &  -0.586 &   0.621 &   4.991 &  -1.015 &   0.700 \\
    4 &   5.613 &  -0.438 &   0.757 &   3.914 &  -0.037 &   0.950 &   4.294 &  -0.283 &   0.625 &   4.027 &  -0.744 &   0.700 \\
    5 &   5.311 &  -0.104 &   0.720 &   3.638 &   0.327 &   0.950 &   4.059 &  -0.003 &   0.601 &   3.857 &  -0.604 &   0.400 \\
    6 &   4.774 &   0.232 &   0.724 &   3.530 &   0.629 &   0.950 &   3.860 &   0.260 &   0.643 &   4.275 &  -0.523 &   0.700 \\
    7 &   4.359 &   0.560 &   0.725 &   3.073 &   1.082 &   0.950 &   3.446 &   0.574 &   0.586 &   3.577 &  -0.036 &   0.579 \\
    8 &   4.087 &   0.835 &   0.743 &   2.450 &   1.475 &   0.950 &   3.071 &   0.787 &   0.776 &   3.011 &   0.374 &   0.700 \\
    9 &   3.682 &   1.160 &   0.751 &   1.832 &   1.708 &   0.950 &   2.560 &   1.002 &   0.300 &   2.698 &   0.776 &   0.491 \\
   10 &   3.316 &   1.414 &   0.838 &   1.300 &   2.132 &   0.950 &   2.741 &   1.190 &   0.585 &   2.466 &   1.011 &   0.582 \\
   11 &   2.744 &   1.703 &   0.835 &   ----- &   ----- &   ----- &   2.244 &   1.485 &   0.662 &   2.235 &   1.276 &   0.404 \\
   12 &   1.618 &   2.249 &   0.720 &   ----- &   ----- &   ----- &   1.482 &   1.734 &   0.758 &   1.761 &   1.580 &   0.590 \\
   13 &   ----- &   ----- &   ----- &   ----- &   ----- &   ----- &   1.235 &   2.068 &   0.800 &   ----- &   ----- &   ----- \\
\hline
 & \multicolumn{3}{c}{NGC 3377} & \multicolumn{3}{c}{NGC 3379} & \multicolumn{3}{c}{NGC 3414} & \multicolumn{3}{c}{NGC 3608} \\
\hline
    1 &   6.042 &  -1.664 &   0.740 &   4.264 &  -1.314 &   0.900 &   5.924 &  -1.704 &   0.801 &   4.387 &  -1.340 &   0.900 \\
    2 &   5.444 &  -0.993 &   0.741 &   4.210 &  -0.771 &   0.900 &   4.990 &  -1.028 &   0.800 &   4.386 &  -0.892 &   0.892 \\
    3 &   4.836 &  -0.604 &   0.469 &   4.182 &  -0.197 &   0.926 &   4.663 &  -0.628 &   0.817 &   4.376 &  -0.566 &   0.890 \\
    4 &   4.837 &  -0.590 &   0.713 &   4.167 &   0.045 &   0.895 &   4.353 &  -0.262 &   0.791 &   4.244 &  -0.237 &   0.834 \\
    5 &   4.173 &  -0.324 &   0.900 &   3.939 &   0.340 &   0.850 &   3.906 &   0.136 &   0.711 &   3.944 &   0.077 &   0.844 \\
    6 &   4.618 &  -0.154 &   0.402 &   3.907 &   0.493 &   0.929 &   3.610 &   0.449 &   0.741 &   3.295 &   0.325 &   0.807 \\
    7 &   4.151 &   0.118 &   0.464 &   3.354 &   0.782 &   0.852 &   3.109 &   0.769 &   0.791 &   3.378 &   0.533 &   0.798 \\
    8 &   3.424 &   0.293 &   0.807 &   3.455 &   0.870 &   0.967 &   2.675 &   0.982 &   0.855 &   3.112 &   0.831 &   0.815 \\
    9 &   3.760 &   0.463 &   0.345 &   2.902 &   1.111 &   0.850 &   2.345 &   1.308 &   0.664 &   2.400 &   1.174 &   0.750 \\
   10 &   3.458 &   0.657 &   0.461 &   2.728 &   1.430 &   0.866 &   2.041 &   1.555 &   0.846 &   2.276 &   1.459 &   0.781 \\
   11 &   3.226 &   0.883 &   0.523 &   2.287 &   1.685 &   0.850 &   1.258 &   1.973 &   0.980 &   1.550 &   1.977 &   0.881 \\
   12 &   2.813 &   1.235 &   0.455 &   1.645 &   2.008 &   0.901 &   ----- &   ----- &   ----- &   ----- &   ----- &   ----- \\
   13 &   2.479 &   1.423 &   0.546 &   1.108 &   2.400 &   0.861 &   ----- &   ----- &   ----- &   ----- &   ----- &   ----- \\
   14 &   1.895 &   1.686 &   0.559 &   ----- &   ----- &   ----- &   ----- &   ----- &   ----- &   ----- &   ----- &   ----- \\
   15 &   1.632 &   1.811 &   0.800 &   ----- &   ----- &   ----- &   ----- &   ----- &   ----- &   ----- &   ----- &   ----- \\
   16 &   1.165 &   2.172 &   0.899 &   ----- &   ----- &   ----- &   ----- &   ----- &   ----- &   ----- &   ----- &   ----- \\
\hline
 & \multicolumn{3}{c}{NGC 4150} & \multicolumn{3}{c}{NGC 4278} & \multicolumn{3}{c}{NGC 4374} & \multicolumn{3}{c}{NGC 4458} \\
\hline
    1 &   5.231 &  -1.421 &   0.990 &   5.147 &  -1.637 &   0.800 &   4.172 &  -1.762 &   0.990 &   4.772 &  -1.100 &   0.250 \\
    2 &   4.528 &  -0.732 &   0.990 &   4.146 &  -0.320 &   0.800 &   4.974 &  -1.762 &   0.990 &   4.970 &  -0.954 &   0.250 \\
    3 &   4.421 &  -0.404 &   0.990 &   4.030 &   0.002 &   0.900 &   4.398 &  -1.093 &   0.990 &   2.748 &  -0.852 &   0.643 \\
    4 &   4.000 &  -0.009 &   0.990 &   3.877 &   0.170 &   0.800 &   3.919 &   0.085 &   0.800 &   4.657 &  -0.655 &   0.533 \\
    5 &   3.676 &   0.326 &   0.701 &   3.784 &   0.423 &   0.842 &   4.080 &   0.412 &   0.800 &   4.233 &  -0.378 &   0.604 \\
    6 &   2.916 &   0.479 &   0.880 &   3.544 &   0.774 &   0.846 &   3.735 &   0.758 &   0.814 &   3.747 &  -0.106 &   0.666 \\
    7 &   2.947 &   0.814 &   0.674 &   3.043 &   1.059 &   0.898 &   3.199 &   1.043 &   0.838 &   3.297 &   0.145 &   0.752 \\
    8 &   2.275 &   1.229 &   0.650 &   2.471 &   1.440 &   0.900 &   2.847 &   1.331 &   0.874 &   3.026 &   0.416 &   0.834 \\
    9 &   2.216 &   1.365 &   0.650 &   1.748 &   1.766 &   0.900 &   2.425 &   1.592 &   0.935 &   2.823 &   0.697 &   0.870 \\
   10 &   1.023 &   1.787 &   0.660 &   1.189 &   2.182 &   0.857 &   1.953 &   1.894 &   0.990 &   2.362 &   1.020 &   0.873 \\
   11 &   ----- &   ----- &   ----- &   ----- &   ----- &   ----- &   1.509 &   2.259 &   0.990 &   2.033 &   1.344 &   0.950 \\
   12 &   ----- &   ----- &   ----- &   ----- &   ----- &   ----- &   ----- &   ----- &   ----- &   0.988 &   1.973 &   0.950 \\
\hline
\end{tabular}
\end{table*}

\begin{table*}
\caption{MGE parameters for the deconvolved $I$-band surface brightness.}
\centering
\begin{tabular}{ccccccccccccc}
\hline
$j$ & $\log I'_j$ & $\log\sigma_j$ & $q'_j$  &  $\log I'_j$ & $\log\sigma_j$ & $q'_j$   & $\log I'_j$ & $\log\sigma_j$ & $q'_j$  & $\log I'_j$ & $\log\sigma_j$ & $q'_j$  \\
    & (L$_{\odot,I}$ pc$^{-2}$) & (arcsec) & & (L$_{\odot,I}$ pc$^{-2}$) & (arcsec) &   & (L$_{\odot,I}$ pc$^{-2}$) & (arcsec) & & (L$_{\odot,I}$ pc$^{-2}$) & (arcsec) &   \\
\hline
 & \multicolumn{3}{c}{NGC 4459} & \multicolumn{3}{c}{NGC 4473} & \multicolumn{3}{c}{NGC 4486} & \multicolumn{3}{c}{NGC 4526} \\
\hline
    1 &   5.592 &  -1.737 &   0.820 &   4.346 &  -1.762 &   0.550 &   6.541 &  -1.762 &   0.950 &   5.463 &  -1.230 &   0.900 \\
    2 &   5.463 &  -1.507 &   0.820 &   3.900 &  -1.137 &   0.545 &   4.738 &  -0.971 &   0.955 &   4.610 &  -0.441 &   0.900 \\
    3 &   4.947 &  -0.928 &   0.817 &   4.277 &  -0.623 &   0.555 &   3.513 &  -0.415 &   0.944 &   4.223 &  -0.067 &   0.900 \\
    4 &   4.560 &  -0.513 &   0.929 &   4.207 &  -0.315 &   0.528 &   3.288 &   0.007 &   0.990 &   4.049 &   0.470 &   0.474 \\
    5 &   4.504 &  -0.197 &   0.949 &   4.134 &  -0.021 &   0.491 &   3.375 &   0.520 &   0.990 &   3.748 &   0.629 &   0.651 \\
    6 &   3.731 &   0.115 &   0.933 &   3.755 &   0.212 &   0.750 &   3.468 &   0.834 &   0.990 &   3.379 &   0.969 &   0.678 \\
    7 &   3.706 &   0.370 &   0.822 &   3.883 &   0.306 &   0.400 &   3.209 &   1.114 &   0.967 &   3.003 &   1.329 &   0.635 \\
    8 &   3.223 &   0.396 &   0.990 &   3.654 &   0.534 &   0.686 &   2.862 &   1.389 &   0.949 &   1.198 &   1.683 &   0.900 \\
    9 &   3.362 &   0.765 &   0.879 &   3.529 &   0.583 &   0.400 &   2.460 &   1.731 &   0.886 &   2.253 &   1.718 &   0.273 \\
   10 &   2.968 &   1.021 &   0.830 &   3.372 &   0.833 &   0.537 &   1.671 &   1.906 &   0.990 &   2.169 &   1.938 &   0.262 \\
   11 &   2.631 &   1.303 &   0.795 &   3.078 &   0.849 &   0.750 &   1.671 &   2.166 &   0.700 &   1.594 &   2.048 &   0.438 \\
   12 &   2.109 &   1.638 &   0.750 &   2.922 &   1.153 &   0.541 &   1.153 &   2.515 &   0.700 &   0.558 &   2.264 &   0.900 \\
   13 &   1.632 &   1.891 &   0.990 &   2.483 &   1.421 &   0.532 &   ----- &   ----- &   ----- &   ----- &   ----- &   ----- \\
   14 &   ----- &   ----- &   ----- &   2.149 &   1.662 &   0.572 &   ----- &   ----- &   ----- &   ----- &   ----- &   ----- \\
   15 &   ----- &   ----- &   ----- &   1.639 &   1.989 &   0.654 &   ----- &   ----- &   ----- &   ----- &   ----- &   ----- \\
\hline
 & \multicolumn{3}{c}{NGC 4450} & \multicolumn{3}{c}{NGC 4552} & \multicolumn{3}{c}{NGC 4621} & \multicolumn{3}{c}{NGC 4660} \\
\hline
    1 &   5.384 &  -1.596 &   0.600 &   5.064 &  -1.762 &   0.990 &   5.797 &  -1.397 &   0.800 &   5.984 &  -1.620 &   0.900 \\
    2 &   4.724 &  -1.038 &   0.600 &   4.282 &  -0.396 &   0.800 &   5.323 &  -1.030 &   0.798 &   5.138 &  -1.071 &   0.900 \\
    3 &   4.181 &  -0.643 &   0.600 &   4.325 &  -0.339 &   0.990 &   4.956 &  -0.701 &   0.773 &   4.853 &  -0.763 &   0.533 \\
    4 &   4.067 &  -0.501 &   0.372 &   4.209 &  -0.057 &   0.957 &   4.557 &  -0.415 &   0.950 &   4.709 &  -0.541 &   0.671 \\
    5 &   3.891 &  -0.234 &   0.600 &   4.153 &   0.223 &   0.933 &   4.200 &  -0.159 &   0.321 &   4.336 &  -0.209 &   0.546 \\
    6 &   3.416 &   0.208 &   0.554 &   3.784 &   0.594 &   0.948 &   4.138 &  -0.062 &   0.893 &   3.997 &  -0.182 &   0.900 \\
    7 &   3.046 &   0.647 &   0.600 &   3.416 &   0.860 &   0.975 &   3.838 &   0.241 &   0.819 &   4.187 &   0.154 &   0.450 \\
    8 &   3.218 &   0.815 &   0.289 &   2.711 &   1.187 &   0.887 &   3.900 &   0.295 &   0.295 &   3.946 &   0.346 &   0.590 \\
    9 &   2.847 &   1.053 &   0.385 &   2.608 &   1.389 &   0.928 &   3.607 &   0.589 &   0.743 &   3.319 &   0.590 &   0.730 \\
   10 &   2.770 &   1.117 &   0.228 &   1.992 &   1.725 &   0.884 &   3.317 &   0.656 &   0.302 &   2.862 &   0.697 &   0.900 \\
   11 &   2.487 &   1.381 &   0.320 &   1.540 &   2.136 &   0.990 &   3.319 &   0.886 &   0.755 &   3.163 &   0.981 &   0.450 \\
   12 &   1.866 &   1.480 &   0.394 &   ----- &   ----- &   ----- &   2.538 &   1.094 &   0.333 &   2.152 &   1.148 &   0.900 \\
   13 &   1.315 &   1.686 &   0.183 &   ----- &   ----- &   ----- &   2.885 &   1.277 &   0.635 &   2.417 &   1.265 &   0.450 \\
   14 &   1.029 &   1.867 &   0.600 &   ----- &   ----- &   ----- &   2.461 &   1.544 &   0.628 &   1.930 &   1.451 &   0.779 \\
   15 &   1.007 &   1.867 &   0.153 &   ----- &   ----- &   ----- &   1.639 &   1.812 &   0.926 &   1.243 &   1.806 &   0.878 \\
   16 &   ----- &   ----- &   ----- &   ----- &   ----- &   ----- &   1.706 &   1.849 &   0.472 &   ----- &   ----- &   ----- \\
   17 &   ----- &   ----- &   ----- &   ----- &   ----- &   ----- &   1.379 &   2.190 &   0.950 &   ----- &   ----- &   ----- \\
\hline
 & \multicolumn{3}{c}{NGC 5813} & \multicolumn{3}{c}{NGC 5845} & \multicolumn{3}{c}{NGC 5846} & \multicolumn{3}{c}{NGC 7457} \\
\hline
    1 &   3.645 &  -0.457 &   0.980 &   5.559 &  -1.762 &   0.800 &   3.647 &  -1.762 &   0.990 &   6.308 &  -1.762 &   0.900 \\
    2 &   3.913 &  -0.262 &   0.980 &   5.229 &  -1.271 &   0.800 &   3.794 &   0.167 &   0.990 &   4.778 &  -1.059 &   0.900 \\
    3 &   3.848 &  -0.026 &   0.900 &   4.688 &  -1.049 &   0.462 &   3.125 &   0.449 &   0.990 &   4.056 &  -0.680 &   0.900 \\
    4 &   3.755 &   0.254 &   0.885 &   4.400 &  -0.567 &   0.800 &   3.334 &   0.725 &   0.990 &   3.702 &  -0.424 &   0.900 \\
    5 &   3.401 &   0.522 &   0.924 &   4.197 &  -0.184 &   0.768 &   2.772 &   1.063 &   0.969 &   3.653 &  -0.132 &   0.696 \\
    6 &   2.962 &   0.810 &   0.909 &   4.232 &   0.015 &   0.263 &   2.625 &   1.289 &   0.974 &   3.451 &   0.172 &   0.703 \\
    7 &   2.516 &   1.024 &   0.882 &   3.862 &   0.063 &   0.708 &   1.946 &   1.526 &   0.850 &   3.174 &   0.495 &   0.613 \\
    8 &   2.264 &   1.336 &   0.784 &   3.710 &   0.237 &   0.668 &   1.984 &   1.777 &   0.924 &   2.541 &   0.812 &   0.528 \\
    9 &   2.180 &   1.623 &   0.721 &   3.629 &   0.459 &   0.800 &   1.431 &   2.197 &   0.850 &   2.535 &   0.939 &   0.655 \\
   10 &   0.706 &   2.063 &   0.980 &   2.632 &   0.664 &   0.538 &   ----- &   ----- &   ----- &   2.520 &   1.256 &   0.555 \\
   11 &   1.467 &   2.063 &   0.720 &   2.909 &   0.820 &   0.564 &   ----- &   ----- &   ----- &   2.236 &   1.628 &   0.517 \\
   12 &   ----- &   ----- &   ----- &   2.127 &   1.206 &   0.708 &   ----- &   ----- &   ----- &   1.369 &   1.958 &   0.774 \\
\hline
\end{tabular}
\label{lastpage}
\end{table*}

In this Appendix (Tables B1--B2) we present the (distance independent) numerical values of the analytically-deconvolved MGE parametrization of the surface brightness of the  galaxies in our sample.\footnote{The values for the MGE parameterization of NGC~2974 were given in \citet{kra05}} The  constant-PA models were obtained by fitting the HST/WFPC2/F814W photometry at small radii and the ground-based MDM photometry (in the same band) at large radii. They provide an accurate description of the $I$-band surface brightness of the galaxies from $R\approx0\farcs01$ to about twice the maximum $\sigma_j$ used in each galaxy (usually corresponding to 5--10 $R_{\rm e}$). The nucleus of NGC~4150 is heavily obscured by dust. For this galaxy we fitted the nuclear MGE model to an HST/NICMOS/F160W ($H$-band) image, which was rescaled to the WFPC2/F814W ($I$-band) one for calibration, neglecting possible $I-H$ color gradients. In other galaxies dust and bright foreground stars were simply excluded from the MGE fit. The matching of the different images and the calibration adopted for the MGE models is described in \refsec{sec:mge}. The deconvolved MGE surface brightness $\Sigma$ is defined as follows:
\begin{equation}
\Sigma(x',y')=\sum_{j=1}^N
I'_j \exp\left[-\frac{1}{2\sigma_j^2}\left(x'^2+\frac{y'^2}{q'^2_j}\right)\right],
  \label{eq:mge_surf}
\end{equation}
where the model is composed by $N$ Gaussian components of dispersion $\sigma_j$, axial ratio $q'_j$ and peak intensity $I'_j$. The coordinates $(x',y')$ are measured on the sky plane, with the $x'$-axis corresponding to the galaxy major axis. The total luminosity of each Gaussian component is given by $L_j=2\pi I'_j\sigma_j^2 q'_j$ \citep[see][for details]{cap02}.

\end{document}